# A quantitative analysis of the 2017 Honduran election and the argument used to defend its outcome

Philip J Gerrish[a,b,1], Benjamin Zepeda[c,d], T Y Okosun[e], Irene A Gerrish[f], and Rosemary Joyce[g]

[a]School of Biology, Georgia Institute of Technology, 310 Ferst Dr., Atlanta GA 30332 USA; [b]Theoretical Biology & Biophysics, Los Alamos National Laboratory, Los Alamos NM 87545 USA; [c]Asociación Coordinadora de Consumidores y Usuarios de Honduras, Tegucigalpa, Honduras; [d]Radio Globo, Edificio Villatoro, Boulevard Morazán, Tegucigalpa, Honduras; [e]Department of Justice Studies, Northeastern Illinois University, Chicago, Illinois 60625 USA; [f]Departments of Economics & Political Science, Hope College, Holland MI 49423 USA; [g]Department of Anthropology, UC Berkely, 232 Kroeber Hall, Berkeley, CA 94720 USA

The Honduran incumbent president and his administration recently declared victory in an election riddled with irregularities and indicators of fraud, not least of which was a numerical anomaly: the primary challenger carried a very significant lead of five percentage points more than half way through the election but was ultimately defeated by the incumbent. The incumbent (Hernandez) offered a plausible explanation for the highly improbable turnaround in the ballots: his popularity is greater in remote areas of the country but votes from remote areas were not counted until later in the election. Here, we mathematically formalize this argument, which we will call the *Hernandez conjecture*, and employ the resulting formulae together with geodemographic data from Honduras to quantitatively assess the conjecture's veracity. We analyze fair-win probabilities assuming that the election outcome was the result of: 1) *geodemography-independent preference* (GIP) only, 2) geodemographics only, and 3) both. When the *departamentos* were analyzed individually, three sparsely-populated *departamentos* (of 18 total) showed small but non-negligible probability of the conjecture's veracity under 2 and 3; however, when the country was analyzed as a whole, the overall probability of the conjecture's veracity was calculated to be less than $10^{-4}$ under a wide range of different assumptions and model variants. Results of our three-pronged analysis, taken together, indicate a negligible probability of a fair win by the incumbent.

Political geography | Electoral dynamics | Mathematical politics | Latin-american studies

In 2009, a constitutional crisis occurred in Honduras, surrounding alleged lifting of a constitutional ban on re-election, that snowballed into a "transfer of power" and the placement of presidential power in the hands of Congress [1, 2]. The United Nations, the Organization of American States (OAS), and various countries refused to acknowledge the de facto government as legitimate, despite the Honduran Supreme Court's claims that the coup was legal. The Supreme Court's assessment was subsequently backed by then US Secretary of State Hillary Clinton's reassuring characterization of these events as little more than a political hiccup [3]. The political lineage instated by this hiccup persists to this day, and the face of that lineage at present is president Juan Orlando Hernandez of the right-wing Partido Nacional (National Party) [4]. This political lineage was a boon to multinationals and was seen as an ally in the Obama-era initiatives to combat corruption in the region, stimulate growth, and reduce the flow of immigrants to the US, especially children [3, 5–11]. The Central American Alliance for Prosperity framework, called "Honduras 2020", was one hopeful face of these initiatives [5, 12]. But the increased presence of multinationals, and the facilitating neoliberal policies behind it, did not translate into a boon for all of the Honduran people [3, 13–16]. Indeed, while a small minority of Hondurans did enjoy a significant boon in personal wealth, the hard evidence indicates that the majority of Honduran people were adversely affected [17–23].

To facilitate the continuation of this political lineage, a series of calculated moves orchestrated by Hernandez over the course of several years set the legislative stage for nullifying the clause in the Honduran constitution prohibiting re-election prior to the 2017 election [1, 3, 17, 18]. The November 2017 election was the first race since the establishment of the constitution during which a president could seek re-election. The President of Honduras is elected by popular vote, whereby the candidate receiving the most votes in a single round of voting is declared the winner.

Hernandez's principal challenger in the 2017 election was Salvador Nasralla, who headed the center-left Alianza movement (Opposition Alliance against the Dictatorship), a coalition of previously established parties consisting of the Partido Libertad y Refundación (LIBRE), Partido Anticorrupción (PAC), and Partido Innovación y Unidad (PINU) parties [24, 25]. As the votes came in on election day, the Supreme Electoral Tribunal (TSE) – which consists primarily of Hernandez loyalists – curiously suspended vote counting for seven

> **Significance Statement**
>
> The recent Honduran presidential election was riddled with irregularities and encumbered with having to explain an extremely improbable turnaround in the ballots more than halfway through the elections. The purported winner nevertheless had a plausible explanation for the turnaround: his popularity is greater in remote areas of the country but votes from remote areas were not counted until later in the election. This argument is difficult to assess verbally, but it is amenable to mathematical formalization and rigorous quantitative assessment. We mathematically formalize the argument and assess its veracity starting as agnostically as possible. We were not able to find a model or set of assumptions for which the probability of the argument's veracity was not very close to zero.

hours and withheld the customary announcement of running totals [26]. During this interval, both of the major candidates declared themselves the winner. When vote counting resumed, the TSE eventually announced their first preliminary results: with 57% of the votes counted, Nasralla carried a seemingly insurmountable lead over Hernandez of five percentage points [3, 27].

At this point, after 57% of the votes had been counted, the TSE halted the vote counting for 36 hours [27, 28]. When the vote counting resumed following this hiatus, Nasralla's lead slowly eroded [28]. When the TSE then took a second hiatus of eight hours, Nasralla decried the election as fraudulent and declared that he would not recognize its outcome [1, 27–30]. Reasons for the interruptions in vote counts included an alleged system failure which, among other things, required the reformatting of computer hard-drives [29].

Hernandez's government issued a 10-day curfew from 6 PM - 6 AM to try and curb protest violence and activity [31–33]. By December 2nd, 7 people had died and 20 were left injured in brutal repression of the protests, a trend that has continued to the present [28, 34, 35]. On December 2nd, the Honduran National Roundtable for Human Rights issued a press release that labeled the government's actions an act of terrorism against its citizens [36].

Hernandez was ultimately declared the winner with 42.95% of the vote to Nasralla's 41.42% [4, 37]. Protests immediately erupted in response to this news [38]. The OAS has officially reported that there have been several irregularities in management of the voting and voting tabulation [37, 39, 40]. In a rare display of bold opposition to the US, secretary general of the OAS Luis Almagro announced that the OAS was calling for a re-election [37]. Hernandez rejected this position, and accused the OAS of being partial to Nasralla.

The spectacular turnaround in the ballots following the interruption in vote counting, while curious indeed, was nevertheless given an entirely plausible explanation by Hernandez and his loyalists: the Partido Nacional enjoys more support in rural areas of the country that are far from major voting centers, and votes cast in these areas required more time to be shipped into the major voting centers and were therefore not counted until later in the election [4, 39, 41]. In what follows, we will refer to this explanation as the *Hernandez conjecture*. Such geographic effects on voting trends are not new [42–44], but some evidence would suggest that they are increasing worldwide [45].

The Economist analyzed the vote tallies as reported by the TSE and compared that information with census data to analyze the validity of the Hernández conjecture [41]. However, The Economist found that explanation implausible, noting that the swing happened in municipalities, which tend to be small and urban, across the country. The only other explanation for the swing in the vote tally would be that paper ballots favored Hernández by 18 percentage points where electronic ballots favored Nasralla by 5 percentage points, but, as The Economist noted, the "odds are that that didn't happen" [41].

Georgetown University professor Irfan Nooruddin performed ballot-data analyses for the Organization of American States and found there to be a sharp swing in the vote-count trend after 68 percent of the votes had been counted [39]. In addition, statistical analyses of the election performed by Honduran engineer Luis Redondo found a qualitative shift in the shape of the voting distributions at around the same time point [46]. Both conclude that these differences are too large to be generated by chance [37, 39, 46, 47].

Here, we complement these previous analyses by carefully scrutinizing the Hernandez conjecture. Our rationale is that the Hernandez conjecture remains a plausible explanation (perhaps the only remaining plausible explanation) for the spectacular ballot turnaround, and its veracity has not formally been assessed. In what follows, we mathematically formalize this conjecture and use the resulting formulae to ask whether the data support it.

## Modeling the *Hernandez conjecture*
### Definitions.

- We will employ the term "half-time" to denote the time at which the vote counts were halted due to an alleged system failure. This point in fact occurred a little more than half way through the elections: the alleged system failure occurred when 57% of the votes were counted.

- We use the word "geodemographics" in a narrow sense referring only to the geographic distribution of the population (i.e., how population density is distributed over the geographic map of Honduras).

- We define $v_H$ and $v_N$ to be the total numbers of votes for Hernandez and Nasralla, respectively, that had been counted by half time, and we define their sum: $v = v_H + v_N$.

- We define $V_H$ and $V_N$ to be total numbers of votes for Hernandez and Nasralla after all votes are counted at the end of the election, respectively, and we define their sum: $V = V_H + V_N$.

- We define independent variable $x$ to be the distance between a voter and the nearest voting center.

- We let $u(x)dx$ denote the number of voters whose distance from the nearest voting center lies within the interval $(x, x + dx)$.

- As defined previously, the *Hernandez conjecture* refers to the argument employed by the incumbent to explain how he won the election fairly despite a significant lead by the opposition at half time. This scenario rests on two fundamental premises:

  1. The *demographics premise*, referring to the incumbent's claim that, as the distance to the nearest voting center increases, people are increasingly likely to vote for the incumbent. This premise is implemented in our model by function $f_H(x)$.

  2. The *vote-count dynamics premise*, referring to the incumbent's claim that, as the distance to the nearest voting center increases, the probability that a person's vote would have been counted by half time decreases. This premise is implemented in our model by function $g(x)$.

Taken together, the foregoing two premises could, in principle, give rise to the spectacular ballot turnaround observed in the 2017 elections that culminated in a win by Hernandez. Generically, this *Hernandez conjecture* is reflected by the simultaneous fulfillment of the conditions $v_N > v_H$ and $V_H > V_N$.

**The model.** Given the foregoing definitions, our model is described in Fig 1 and gives rise immediately to vote counts at half-time:

$$v_i = \int_0^{x_M} f_i(x)g(x)u(x)dx , \quad [1]$$

where $i \in \{H, N\}$, $H$ denoting Hernandez and $N$ Nasralla; and after all votes had been counted:

$$V_i = \int_0^{x_M} f_i(x)u(x)dx \quad [2]$$

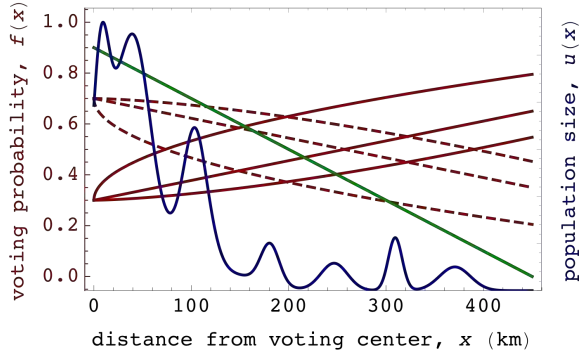

**Fig. 1.** Schematic of our mathematical model of the *Hernandez conjecture*. Independent variable $x$ (horizontal axis) is the distance between a voter and the nearest major voting center. Solid red curves plot probabilities of voting for Hernandez, $f_H(x)$ (three different plausible curves); these probabilities are some increasing function of $x$. Dashed red curves plot probabilities of voting for Nasralla, $f_N(x)$ (three different plausible curves); these probabilities decrease in $x$. The blue curve represents a plausible plot of voter-population size $u(x)$ as a function of distance from the nearest major voting center. The green line plots a plausible function $g(x)$, defined as the probability that a vote cast at distance $x$ from the nearest voting center would have been bussed in and counted by half time. In most of our analyses, we make the assumption that $g(x_M) = 0$, where $x_M$ is the maximum distance between a voter and the nearest voting center. This assumption facilitates the analysis, making it non-parametric, and it also is a conservative assumption in the sense that it favors the *Hernandez conjecture*.

## Summary of results

**Our three-pronged approach.** To glean as much information from the numbers as we can while relying on the fewest possible assumptions, we take a three-pronged approach that begins as agnostically as possible. Throughout our analyses, where assumptions are necessary, we consistently opt for the most conservative assumptions, i.e., assumptions that most strongly favor the Hernandez conjecture. The three prongs of our analyses and results therein are summarized here:

1. **The "no-geodemographics" extreme.** In this extreme, we treat the vote counts as if they were coming from a single, completely homogeneous, population; in this population, the rate at which votes are counted is the same throughout. Standard statistical approaches are readily applied to such a population. Given ballot data from the first half of the election, we compute the probability of a fair win by Hernandez to be:

$$\int \mathbb{P}\{V_H > V_N | V_N\} \mathbb{P}\{V_N\} dV_N = \int (1 - \Phi_Y) d\Phi_Z \quad [3]$$

where $Y = (V_N/V - p)/\sqrt{p(1-p)/v}$ and $Z = (V_N/V - q)/\sqrt{q(1-q)/v}$, with $p = v_H/v$, $q = v_N/v$. As commonly denoted, $\Phi$ is the *cdf* of the unit normal distribution. Computed under this "no-geodemographics" assumption, the probability of a fair win by Hernandez is effectively zero ($\approx 10^{-1770}$).

2. **The "all-geodemographics" extreme.** In this extreme, we assume that a person's tendency to vote for one candidate or the other depends *exclusively* on where that person lives: averaged over the entire terrain in question (an average taken over space), the net preference for one candidate over the other is zero. The outcome of the election, in this extreme, is a product of one thing only, namely, how the population is distributed throughout the region under scrutiny. Assuming those farthest from voting centers had zero probability of being counted in the first half of the election (an assumption favoring the Hernandez conjecture), and implementing linear functions for the spatial dependencies, we find that the relevant conditions reduce to compact expressions that very conveniently have no parameters to be estimated. The condition for a fair win by Hernandez in this extreme is:

$$\mathbb{E}(X - \tfrac{1}{2}x_M) > 0 , \quad [4]$$

where random variable $X$ is the distance between a voter and the nearest voting center, and $x_M = \max(X)$. The condition for a half-time lead by Nasralla in this extreme is:

$$\mathbb{E}[(X - \tfrac{1}{2}x_M)(X - x_M)] > 0 \quad [5]$$

The condition for a turnaround in the ballots such as the one observed in Honduras is thus a condition in which Eq. (4) and Eq. (5) are met simultaneously. The simultaneous fulfillment of Eq. (4) and Eq. (5) thus constitutes a mathematical formalization of the Hernandez conjecture in the all-geodemographics extreme. We note that Eq. (4) and Eq. (5) do not depend in any way on ballot data; they depend only on the geodemographics of Honduras. These developments readily provide a non-parametric means to assess whether the geodemographics of Honduras alone are sufficient to: A) promote a win by Hernandez Eq. (4), and/or B) support the Hernandez conjecture (Eq. (4) and Eq. (5)). Using geodemographic data from Honduras obtained from Wolfram databases [48], we tried several different ways to distribute voting centers throughout the departments and throughout the country; in addition, we explored logarithmic and power-law spatial dependencies in addition to linear; however, we were unable to find a configuration for which the probability of the Hernandez conjecture was not essentially zero (in many cases too small to be computed by standard numerical integration routines). When the different *departamentos* were treated individually, we again found the Hernandez conjecture to be essentially zero in all *departamentos* except for one, namely, Gracias a Dios which had non-negligible upper-bounds under some configurations (see Figs 6, 7, and 8).

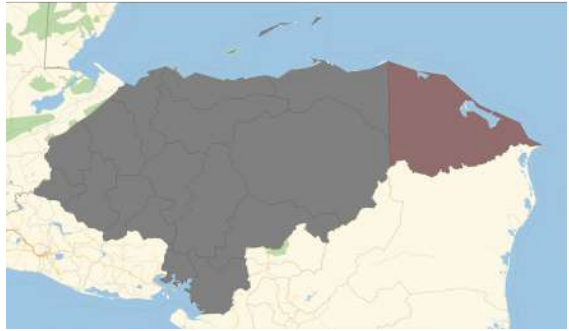

**Fig. 2.** Probability of the Hernandez conjecture by *departamento*, in the "all-geodemographics" extreme. Colors indicate the probability of the conjecture, computed as the fraction of bootstrapped data sets for which Eq. (4) and Eq. (5) are met simultaneously. Here, we assumed one major voting center per departamento. Probabilities are essentially zero (gray) in 17 departamentos but were non-negligible in one departamento, namely Gracias a Dios. A look at Fig 5 shows these this departamento to be among the more sparsely populated.

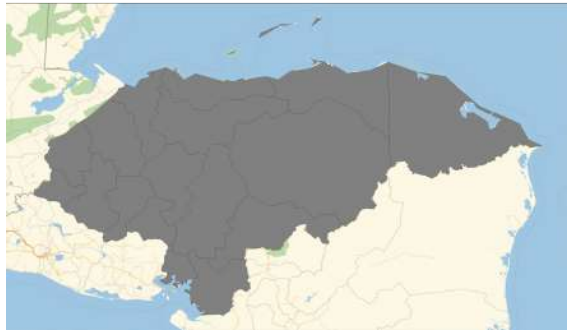

**Fig. 3.** Probability of the Hernandez conjecture by *departamento*, in the "all-geodemographics" extreme. Same as 2 but assuming 3 major voting centers per departamento. Probabilities are essentially zero in all 18 departamentos.

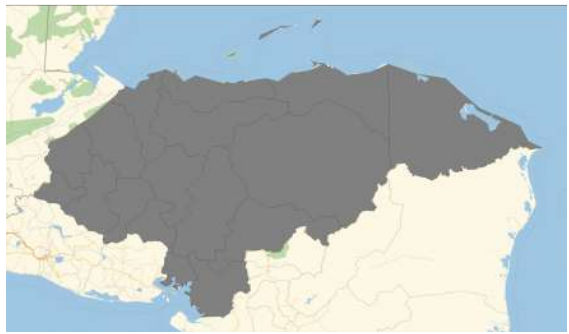

**Fig. 4.** Probability of the Hernandez conjecture by *departamento*, in the "all-geodemographics" extreme. Same as 2 but assuming 5 major voting centers per departamento. Probabilities are essentially zero in all 18 departamentos.

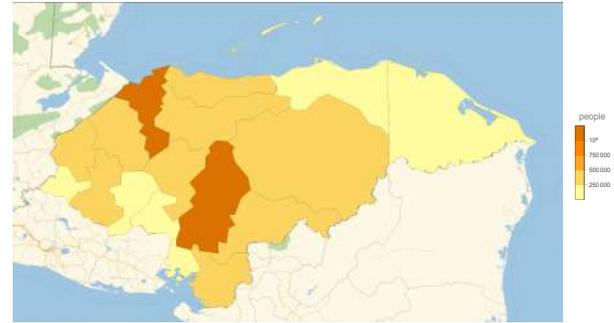

**Fig. 5.** Population size by *departamento*.

3. **The "partial-geodemographics" intermediate.** Having explored the two extreme cases outlined above, we next move to middle ground in which outcomes are partially due to geodemographics and partially to an overall *geodemography-independent preference*, or GIP. More specifically, GIP refers to the component of a voter's preference that is independent of where that voter lives. Under the Hernandez conjecture, the GIP is constrained by two opposing requirements: on one hand, it should be large enough to favor the candidate in question (in this case Hernandez), while on the other hand it cannot be too large or the observed ballot turnaround cannot be explained. We show that, under linear dependencies and our conservative assumption outlined in prong 2 above, the numerical value of GIP – to which we will assign the variable $\epsilon$ – is constrained by the Hernandez conjecture to reside within a critical interval $\epsilon \in (\epsilon_l, \epsilon_u)$. We show that, under the conservative assumptions outlined above (assumptions that favor the Hernandez conjecture), $\epsilon \in (\epsilon_l, \epsilon_u)$ when:

$$\frac{\mathbb{E}[(X - \frac{1}{2}x_M)(X - x_M)]}{\mathbb{E}(X - x_M)} - \mathbb{E}(X - \frac{1}{2}x_M) \\ < \frac{c}{m}\left(\frac{v_H - v_N}{v_H + v_N}\right) < 0, \quad [6]$$

where $c$ and $m$ are parameters of $h(x)$ (constant and gradient, respectively). We note that the left-hand side of the inequality is guaranteed to be negative by Jensen's inequality. Parameters $c$ and $m$ cannot be assumed to be independent, and the distribution of their ratio is thus computed by repeated fitting of $h(x)$ to bootstrapped data. The probability of the Hernandez conjecture is then computed simply as the fraction of bootstrapped values of $c/m$ that put the middle term in Eq. (6) within the prescribed bounds. Figures 6, 7, and 8 show probabilities that the geodemographics of Honduras together with ballot data support the Hernandez conjecture; in addition, we explored logarithmic and power-law spatial dependencies in addition to linear; these probabilities are only non-negligibly greater than zero in three of the more sparsely-populated *departamentos*. When all of the data are analyzed together, the country-wide probability of the Hernandez conjecture is less that 0.0001 for all configurations analyzed.

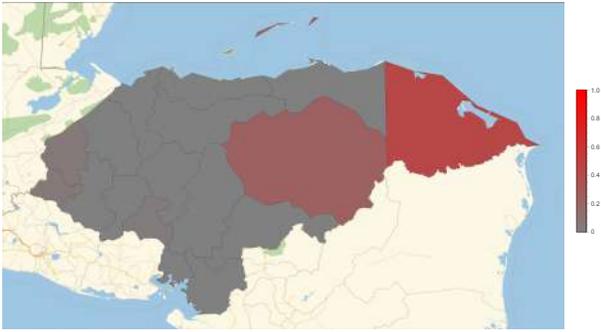

**Fig. 6.** Probability of the Hernandez conjecture by $departamento$. Colors indicate the values of $\mathbb{P}\{\hat{\epsilon} \in (\epsilon_l, \epsilon_u)\}$, computed by Eq. (6), which is the probability of the Hernandez conjecture. Here, we assumed one major voting center per departamento. Probabilities are very close to zero (gray) in 15 departamentos but are non-negligible in three departamentos, namely, Gracias a Dios, Islas de la Bahia, and Olancho. A look at Fig 5 shows these three departamentos to be among the more sparsely populated.

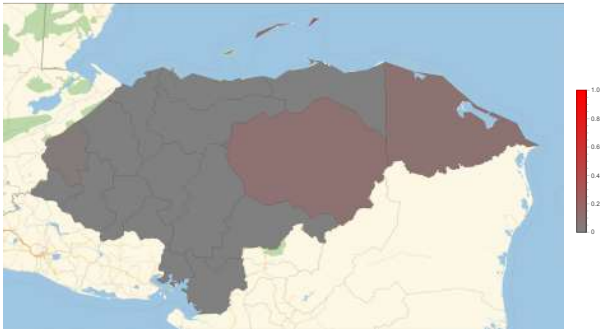

**Fig. 7.** Probability of the Hernandez conjecture by $departamento$. Same as 6 but here we assume 3 major voting centers per departamento. Probabilities are very close to zero (gray) in 15 departamentos but are non-negligible in three departamentos, namely, Gracias a Dios, Islas de la Bahia, and Olancho. A look at Fig 5 shows these three departamentos to be among the more sparsely populated.

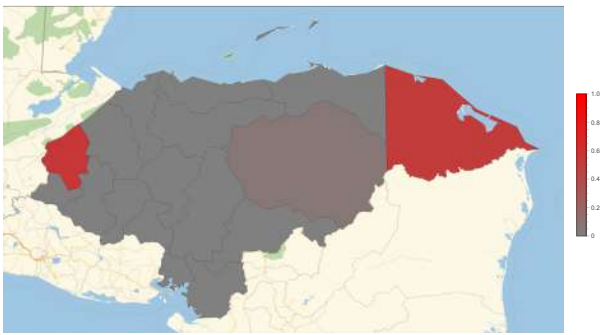

**Fig. 8.** Probability of the Hernandez conjecture by $departamento$. Same as 6 but here we assume 5 major voting centers per departamento. Probabilities are very close to zero (gray) in 15 departamentos but are non-negligible in three departamentos, namely, Gracias a Dios, Islas de la Bahia, and Olancho. A look at Fig 5 shows these three departamentos to be among the more sparsely populated.

## Conclusions

We have rigorously assessed the *Hernandez conjecture* and found no configuration of ballot data, geodemographic data, or both, that supports it. Our analyses have been both: 1) as agnostic as possible eliminating dependence on assumptions and ballot data where possible, and 2) as generous as possible to the conjecture, intentionally giving it the upper hand wherever assumptions were required. And still, our calculations indicate the probability of the conjecture's veracity to be essentially zero under a wide range of different conditions. We comment that, were the present study conducted within the rigorous world of scientific research, the weight of the evidence presented here should be sufficient to robustly support rejection of the conjecture.

Our findings corroborate previous statistical analyses in their assessment of the elections [39, 41, 46]. And given the absence of a plausible alternative to the Hernandez conjecture, it follows that our analysis leaves little room for any remaining confidence in the fairness of this election.

## Discussion: the broader context

Our conclusions, while concerning, could be seen as a microcosm of a global trend – one that shows increasing susceptibility to populism and its unfortunate ripple effects, namely, increasingly callous foreign policies as well as autocratic and dictatorial governance. Indeed, the 2017 Honduran elections and the reshaping of the Honduran Constitution is a window into this trend. And Honduras is not alone: for example, Honduras is almost an exact replication of the scenario in Cote d'Ivoire (Ivory Coast) in 2010, where the incumbent Laurent Gbagbo refused to relinquish power to Allasane Ouattara who clearly won the popular vote [49].

Historical evidence overwhelmingly supports the notion that citizens and governing systems are truly functional only within a context of fairness [50, 51]. Implicit in such fairness is unfettered democracy, a respect for citizens' rights and participation. The universalizability principle put forth by Benn [52] suggests there is something to be gained from the extrapolation of political microcosms such as the one studied here to the global community.

Finally, the response of the US was a familiar one [53, 54]. Soon after the first and tentative announcement of Hernandez's victory, and despite clear indications of fraud, the incumbent administration was cleared to receive its share of $644 million in continued US aid – military and other [55–60]. This award, poignantly, is strongly conditional on a positive assessment of human rights and progress combating corruption and inequality [19, 61–65]. The analysis we have presented here discredits the *Hernandez conjecture*, which was very probably the incumbent's only remaining plausible defense against accusations of fraud.

## Acknowledgements

Special thanks to John Gerrish for helpful discussions and suggestions. Many thanks to Ing. Luis Redondo for sending and referring us to essential data sets. This work was supported in part by a CNRS visiting scholar grant (to PG), and Hope College scholarships (to IG).

# Supplementary Information for

## A quantitative analysis of the 2017 Honduran election and the argument used to defend its outcome

**Philip J Gerrish**[1,2]**, Benjamin Zepeda**[3,4]**, T Y Okosun**[5]**, Irene A Gerrish**[6] **& Rosemary Joyce**[7]

**Philip J Gerrish**
E-mail: pgerrish@gatech.edu

**This PDF file includes:**

> Supplementary text
> Figs. S1 to S16
> References for SI reference citations

**Supporting Information Text**

**Model overview.**

**Definitions:**

- We will employ the term "half-time" to denote the time at which the vote counts were halted due to an alleged system failure. This point in fact occurred a little more than half way through the elections: the alleged system failure occurred when 57% of the votes were counted.

- We use the word "geodemographics" in a narrow sense referring only to the geographic distribution of the population (i.e., how population density is distributed over the geographic map of Honduras).

- We define $v_H$ and $v_N$ to be the total numbers of votes for Hernandez and Nasralla, respectively, that had been counted by half time, and we define their sum: $v = v_H + v_N$.

- We define $V_H$ and $V_N$ to be total numbers of votes for Hernandez and Nasralla after all votes are counted at the end of the election, respectively, and we define their sum: $V = V_H + V_N$.

- We define independent variable $x$ to be the distance between a voter and the nearest voting center.

- We let $u(x)dx$ denote the number of voters whose distance from the nearest voting center lies within the interval $(x, x + dx)$.

- As defined previously, the *Hernandez conjecture* refers to the argument employed by the incumbent to explain how he won the election fairly despite a significant lead by the opposition at half time. This scenario rests on two fundamental premises:

  1. The *demographics premise*, referring to the incumbent's claim that, as the distance to the nearest voting center increases, people are increasingly likely to vote for the incumbent. This premise is implemented in our model by function $f_H(x)$.

  2. The *vote-count dynamics premise*, referring to the incumbent's claim that, as the distance to the nearest voting center increases, the probability that a person's vote would have been counted by half time decreases. This premise is implemented in our model by function $g(x)$.

  Taken together, the foregoing two premises could, in principle, give rise to the spectacular ballot turnaround observed in the 2017 elections that culminated in a win by Hernandez. Generically, this *Hernandez conjecture* is reflected by the simultaneous fulfillment of the conditions $v_N > v_H$ and $V_H > V_N$.

- We define $\omega(x) = ku(x)dx$, where $k = \left(\int_0^{x_M} u(x)dx\right)^{-1}$ and $x_M$ is the maximum distance from an individual to the nearest voting center, i.e., $\omega(x)$ is the probability density of population as a function of distance from nearest voting center, $x$.

- We define:
$$f_H(x) = \mathbb{P}\{\text{vote H} \mid x\}$$
$$f_N(x) = \mathbb{P}\{\text{vote N} \mid x\},$$

  i.e., the probabilities that a person will vote for Hernandez and Nasralla, respectively, given that the distance between the person and the nearest voting center is $x$.

- We define:
$$g(x) = \mathbb{P}\{\text{vote counted at half time} \mid x\},$$

  i.e., the probability that a person's vote is counted by half time, given that the person's distance from the nearest voting center is $x$.

- Given the foregoing definitions, the fraction of votes going to Hernandez and Nasralla at half time is, respectively,
$$v_H = \mathbb{E}[f_H(X)g(X)]$$
and
$$v_N = \mathbb{E}[f_N(X)g(X)]$$

  where $X \sim \omega(x)$.

 Philip J Gerrish[1,2], Benjamin Zepeda[3,4], T Y Okosun[5], Irene A Gerrish[6] & Rosemary Joyce[7]

- Likewise, the fraction of votes going to Hernandez and Nasralla when all votes are counted is:
$$V_H = \mathbb{E}[f_H(X)]$$
and
$$V_N = \mathbb{E}[f_N(X)]$$
where $X \sim \omega(x)$.

- *Modeling the effects of geodemographics.* We define function $h(x)$ with the properties,
$$\int_0^{x_M} h(x)dx = 0 \quad \text{and} \quad \int_0^{x_M} h'(x)dx > 0$$
This function $h(x)$ we will call the *geodemography component*. The second of the two foregoing conditions guarantees that $h(x)$ increases more than it decreases such that there is a net upward displacement in $h(x)$ as $x$ increases ($h(x_M) - h(0) > 0$) – consistent with the *Hernandez conjecture*.

    – Case 1: The "all-geodemographics" extreme:
    $$f_H(x) = c + h(x)$$
    $$f_N(x) = c - h(x),$$
    for some constant $c$ which delineates an axis of reflection. We call this the "all demographics" extreme, because there is no effect of overall voter preference: outcomes result exclusively from demographic and vote-count dynamics effects. This fact derives from the observation that, given the above definitions we have:
    $$\int_0^{x_M} f_H(x)dx = \int_0^{x_M} f_N(x)dx$$

    – Case 2: The "partial-geodemographics intermediate". This case allows for outcomes that are partly due to demographics and vote-count dynamics, and partly due to voter preference that is independent of where a voter lives. We define an index of overall geodemography-independent preference, $\epsilon$, as follows:
    $$f_H(x) = c + \epsilon + h(x)$$
    $$f_N(x) = c - \epsilon - h(x)$$
    When overall voter preference favors Nasralla, $\epsilon$ is negative.

**Assumption:** The single assumption made by all of our analyses is that people's tendency to vote for a given party was not affect by the system shut-down that occurred at half-time.

**Objectives:** Our approach is to start at two extremes, both of which require a reduced set of assumptions and data, and work our way toward middle ground.

1. Assess the likelihood of the *Hernandez conjecture* under the assumption of a "well-mixed" population (the "no-geodemographics" extreme). No geodemographic data are required for this analysis.
2. Assess whether the geodemography of Honduras could alone support the *Hernandez conjecture* (the "all-demographics" extreme). No ballot data are required for this analysis.
3. Compute the probability of the *Hernandez conjecture* given both the geodemography of Honduras and the vote counts at half-time. Both ballot data and geodemographic data are used in this analysis.

**Objective 1: Fair win probability in the "no geodemographics" extreme..** In this first analysis, we will ignore the demographic argument put forth by Hernandez loyalists, with the purpose of gaining insight into the magnitude of the probabilities under a completely agnostic model.

Let $\hat{p}$ and $\hat{q}$ denote estimators of the probabilities that an individual votes for Hernandez and Nasralla, respectively. These are Binomial sampling estimators with associated normal distributions:
$$\hat{p} \sim \mathcal{N}(p, \sigma_p^2) \quad \text{and} \quad \hat{q} \sim \mathcal{N}(q, \sigma_q^2),$$
where $\sigma_p^2 = p(1-p)/N$, $\sigma_q^2 = q(1-q)/N$, $p = n_H/N$, $q = n_N/N$, $N$ = total number of votes cast, $n_H$ = number of votes cast for Hernandez, and $n_N$ = number of votes cast for Nasralla. The probability of a win by Hernandez is thus:
$$\mathbb{E}_X[(1 - \Phi_X(p, \sigma_p^2))] \quad \text{where} \quad X \sim \mathcal{N}_X(q, \sigma_q^2),$$
and $\Phi$ denotes the cumulative normal distribution. Simply taking the numbers given just before the purported "system failure", and assuming: 1) no effects of changing demographics, and 2) no tendency for people to have a change of heart due to the "system failure" and vote for Hernandez, we find that the probability of an ultimate win by Hernandez to be roughly $10^{-1770}$.

**Objective 2: Fair win possibility in the "all geodemographics" extreme..** Our subsequent analyses give as much credence as possible to the Hernandez "victory scenario", in order to determine the likelihood of a win by Hernandez when he is favored by a "best-case-scenario".

- Nasralla's lead at half time may be written mathematically as $v_N > v_H$, or:

$$\mathbb{E}[f_N(X)g(X)] > \mathbb{E}[f_H(X)g(X)]$$

which may be rewritten as:

$$\mathbb{E}[(c - h(X))g(X)] > \mathbb{E}[(c + h(X))g(X)]$$

which reduces to the condition:

$$\boxed{\mathbb{E}[g(X)h(X)] < 0} \quad [1]$$

- Hernandez's overall lead may be written mathematically as $V_H > V_N$, or:

$$\mathbb{E}[f_H(X)] > \mathbb{E}[f_N(X)]$$

which may be written as:

$$\mathbb{E}[c + h(X)] > \mathbb{E}[c - h(X)]$$

which reduces to the condition:

$$\boxed{\mathbb{E}[h(X)] > 0} \quad [2]$$

Here, we would like to look qualitatively for scenarios in which vote counts at half time could put Nasralla ahead:

$$\mathbb{E}[g(X)h(X)] < 0$$

while hiding the fact the Hernandez was in fact winning:

$$\mathbb{E}[h(X)] > 0$$

**Linear model.** Here, we assume that the functions $g(x)$, $f_H(x)$, and $f_N(x)$ are linear. For simplicity, we will assume $f_H(x)$ and $f_N(x)$ are mirror images of one another. These linear functions are thus given by:

$$f_H(x) = a + mx$$
$$f_N(x) = b - mx$$

where $a$ and $b$ are lower and upper bounds, and $m = (b-a)/x_M$. We further assume that $g(x)$ is a linearly decreasing function of distance, given by:

$$g(x) = c - dx \ .$$

**Condition for fair win by Hernandez.** Our first observations are that, all else being equal, Hernandez wins legitimately ($V_H > V_N$) when:

$$\int_0^{x_M} f_H(x)u(x)dx > \int_0^{x_M} f_N(x)u(x)dx$$

or:

$$\int_0^{x_M} (a + mx)u(x)dx > \int_0^{x_M} (b - mx)u(x)dx$$

Rearranging gives:

$$2m \int_0^{x_M} xu(x)dx > \int_0^{x_M} (b - a)u(x)dx$$

or:

$$2\left(\frac{b-a}{x_M}\right)\int_0^{x_M} xu(x)dx > (b - a)\int_0^{x_M} u(x)dx$$

Noting that $\int_0^{x_M} u(x)dx = N$ and $\int_0^{x_M} xu(x)dx = N\mathbb{E}(X)$, where $\mathbb{E}(X)$ is the mean distance between individuals and the nearest voting center. Rearranging, we have that under a linear model, the condition for a fair win by Hernandez (Eq. (2)) reduces to:

$$\mathbb{E}(X - \frac{1}{2}x_M) > 0 \quad [3]$$

where $X \sim \omega(x)$, and recalling that $\omega(x) = u(x)/N$. This very simple and intuitive result provides an easy test to see whether Honduran geodemography alone could give rise to a win by Hernandez. The only assumption made is that voters that live further from voting centers are more likely to vote for Hernandez; no model parameters are required. We apply Eq. (2) to Honduran geodemographic data to compute the probability that geodemography *by itself* supports a fair win by the incumbent.

 Philip J Gerrish[1,2], Benjamin Zepeda[3,4], T Y Okosun[5], Irene A Gerrish[6] & Rosemary Joyce[7]

**Condition for half-time lead by Nasralla.** Nasralla leads at half time ($v_H > v_N$) when:

$$\int_0^{x_M} f_N(x)g(x)u(x)dx > \int_0^{x_M} f_H(x)g(x)u(x)dx$$

or:

$$\int_0^{x_M} (b-mx)(c-dx)u(x)dx > \int_0^{x_M} (a+mx)(c-dx)u(x)dx$$

Again, we have that $m = (b-a)/x_M$. We now make a conservative assumption favoring the Hernandez conjecture: we assume that a vote cast at the farthest point from the voting center will have been shipped in and counted by half time with probability zero. This assumption gives us $d = c/x_M$. Rearranging gives:

$$\mathbb{E}[(X - \tfrac{1}{2}x_M)(X - x_M)] > 0 \qquad [4]$$

**The "all-geodemographics" condition under which the opposition's half time lead could obfuscate an underlying lead by the incumbent.** This condition is met when conditions Eqs Eq. (3) and Eq. (4) are met simultaneously. Applying these conditions, Honduran geodemographic data alone (no ballot data necessary) support the victory scenario with probability effectively equal to zero. We reiterate, however, that this is for the extreme in which there is no geodemography-independent preference (GIP) one way or the other, i.e., geodemography in this case is assumed to be the only factor influencing the outcome.

**Objective 3: Inferring GIP signal based on half-time vote counts..**

*General model.*

- The claim we would like to evaluate is that:
$$V_H > V_N$$
  despite the fact that:
$$v_H = 0.41 \text{ and } v_N = 0.46$$

- Conditions for a win by Hernandez are given when $V_H > V_N$, as defined above.

  – Case 1: Under "all demographics" assumption, this condition is met when:
$$\int_0^{x_M} (c + h(x))u(x)dx > \int_0^{x_M} (c - h(x))u(x)dx$$

  which reduces to the condition:
$$\int_0^{x_M} h(x)u(x)dx > 0$$

  We note that total population size is $N = \int_0^{x_M} u(x)dx$, and we define a new function $\omega(x) = u(x)/N$ which has the property $\int_0^{x_M} \omega(x)dx = 1$. We can now rewrite the above condition as:
$$\int_0^{x_M} h(x)\omega(x)dx > 0,$$

  or equivalently, Hernandez wins under the "all demographics" assumption when:
$$\boxed{\mathbb{E}[h(X)] > 0 \;,\;\; \text{where} \;\; X \sim \omega(x)} \qquad [5]$$

  – Case 2: When there is an overall voter preference, we have:
$$\int_0^{x_M} (c + \epsilon + h(x))u(x)dx > \int_0^{x_M} (c - \epsilon - h(x))u(x)dx \;,$$

  which, when the above definitions are applied reduces to the general condition for a win by Hernandez:
$$\boxed{\mathbb{E}[h(X)] > -\epsilon \;,\;\; \text{where} \;\; X \sim \omega(x)} \qquad [6]$$

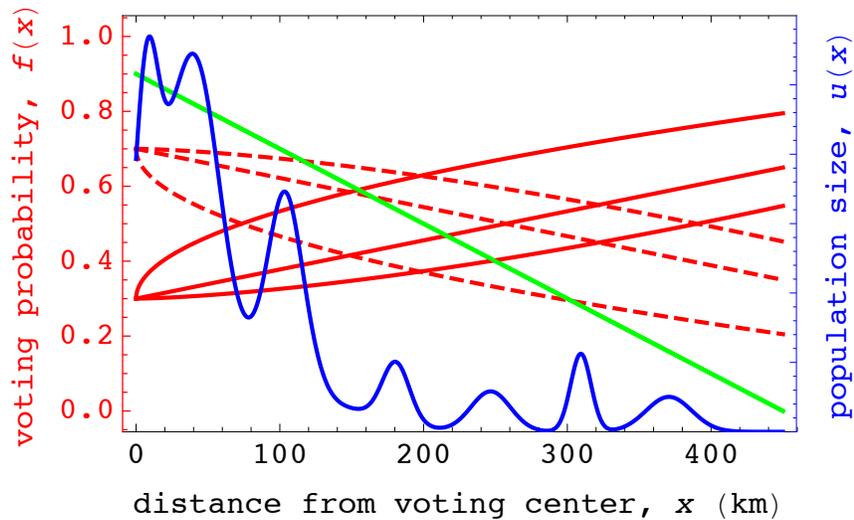

**Fig. S1.** Schematic of voting probabilities and population size as a function of distance from the nearest major voting center. Solid red curves plot probabilities of voting for Hernandez, $f_H(x)$ (three different plausible curves); these probabilities increase with distance from major voting centers to reflect the victory-scenario argument of Hernandez loyalists. Dashed red curves plot probabilities of voting for Nasralla, $f_N(x)$ (three different plausible curves); these probabilities decrease with distance from major voting centers, again to reflect the victory-scenario argument. The blue curve represents a plausible plot of population size as a function of distance from the nearest major voting center, $u(x)$. The green line plots plausible probabilities that a vote cast at distance $x$ from the voting center would have been bussed in to the voting center and counted half way through the election (when the purported system failure occurred); this function we will denote as $g(x)$.

**Philip J Gerrish**[1,2], **Benjamin Zepeda**[3,4], **T Y Okosun**[5], **Irene A Gerrish**[6] & **Rosemary Joyce**[7]

***Vote counts at "half time".*** The vote counts at "half time" when the alleged system failure occurred are given by:

$$v_H = \int_0^{x_M} f_H(x)g(x)u(x)dx = \int_0^{x_M} (c + \epsilon + h(x))g(x)u(x)dx$$

$$v_N = \int_0^{x_M} f_N(x)g(x)u(x)dx = \int_0^{x_M} (c - \epsilon - h(x))g(x)u(x)dx$$

At half time, Nasralla was winning by known factor $r > 1$, supplying us with the following information:

$$r\int_0^{x_M} (c + \epsilon + h(x))g(x)u(x)dx = \int_0^{x_M} (c - \epsilon - h(x))g(x)u(x)dx$$

or, rearranging:

$$\mathbb{E}[h(X)g(X)] = [\frac{1-r}{1+r}c - \epsilon]\mathbb{E}[g(X)]$$

which gives:

$$\epsilon = \frac{1-r}{1+r}c - \frac{\mathbb{E}[h(X)g(X)]}{\mathbb{E}[g(X)]} \qquad [7]$$

or, rewriting, vote counts $v_H$ and $v_N$ at half-time reflect an overall geodemographics-independent preference (GIP) for Hernandez of:

$$\epsilon = c\left(\frac{v_H - v_N}{v_H + v_N}\right) - \frac{\mathbb{E}[h(X)g(X)]}{\mathbb{E}[g(X)]} \qquad [8]$$

When this quantity is negative, there is overall GIP for Nasralla. To compute an overall fair win by Hernandez given half-time ballot data, therefore, condition Eq. (6) must hold, where $\epsilon$ is defined by Eq. (9), giving the condition:

$$\mathbb{E}[h(X)] > \frac{\mathbb{E}[h(X)g(X)]}{\mathbb{E}[g(X)]} - c\left(\frac{v_H - v_N}{v_H + v_N}\right) \qquad [9]$$

or generally:

$$\boxed{c\left(\frac{v_H - v_N}{v_H + v_N}\right)\mathbb{E}[g(X)] \;>\; \mathbb{E}[h(X)g(X)] - \mathbb{E}[h(X)]\mathbb{E}[g(X)]} \qquad [10]$$

Inserting linear functions for $g(X)$ and $h(X)$, and again employing the conservative assumption $g(x_M) = 0$, gives the condition:

$$\frac{\mathbb{E}[(X - \frac{1}{2}x_M)(X - x_M)]}{\mathbb{E}(X - x_M)} \;-\; \mathbb{E}(X - \frac{1}{2}x_M) \;<\; \frac{c}{m}\left(\frac{v_H - v_N}{v_H + v_N}\right) \;<\; 0 \;, \qquad [11]$$

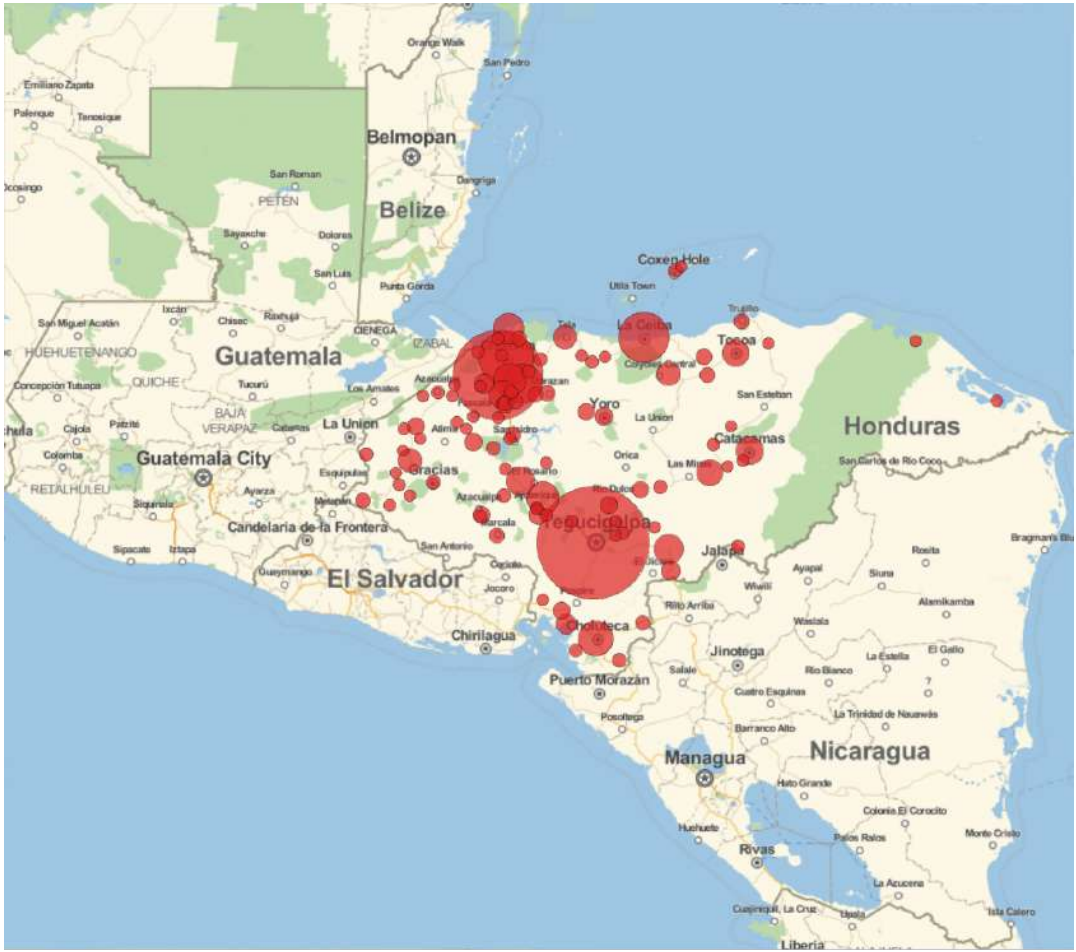

**Fig. S2.** Demographics of Honduras. Red circles plot the larger cities and towns of Honduras. Circle size is an indicator of population size.

   Philip J Gerrish[1,2], Benjamin Zepeda[3,4], T Y Okosun[5], Irene A Gerrish[6] & Rosemary Joyce[7]

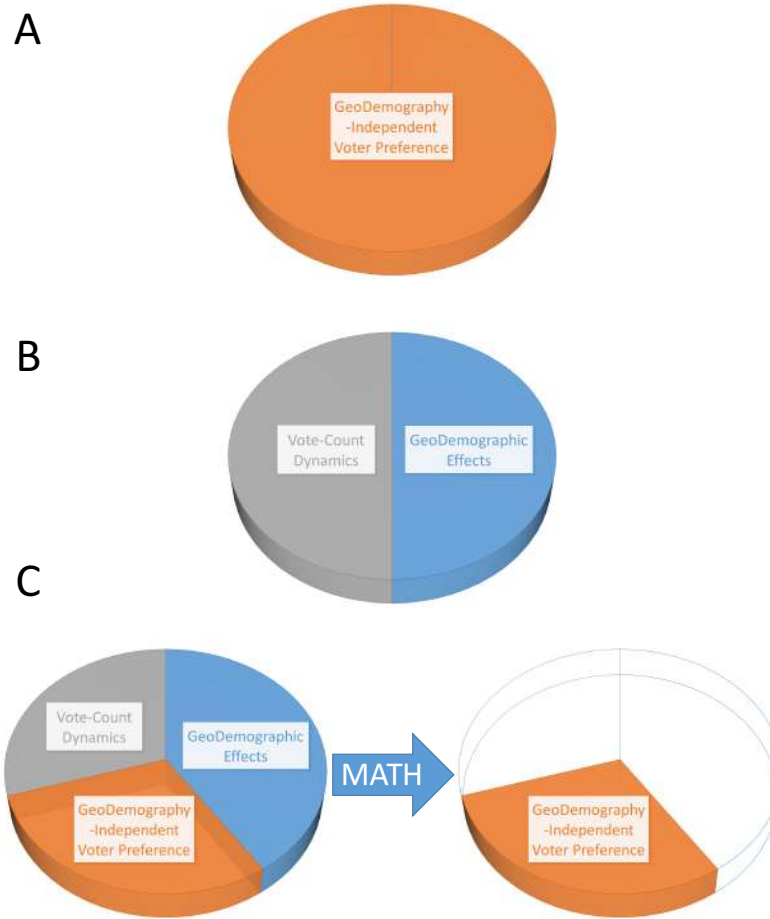

**Fig. S3. Three-pronged approach**. **A**, Our first approach was to ignore geodemography, i.e., to assume that one's tendency to vote one way or the other was $completely\ independent$ of where one lives. This straightforward calculation revealed that the probability of a fair win by Hernandez is zero, suggesting that a fair win would require some artifactual explanation. An entirely plausible such explanation was put forth by the Hernandez camp and this we refer to as the $Hernandez\ conjecture$. **B**, Our second approach was to assume that geodemography was the only factor affecting the outcome, i.e., to assume that one's tendency to vote one way or the other was $entirely\ dependent$ on where one lives, so that if one took an average over $space$ there would be no net preference for one candidate over the other. The geodemographics effect in this case has two components: the component of how the population is distributed over the country of Honduras (geodemographics effects), and the vote-count dynamics that resulted in votes closer to voting centers being more likely to have been counted by half-time. These calculations again revealed that the probability of a fair win by Hernandez is zero when the country is analyzed as a whole but when each $departamento$ is analyzed separately a non-negligible probability was computed for two sparsely-populated $departamentos$. **C**, Our third approach was to assume there were essentially three components to one's votes: vote-count dynamics, geodemographics, and geodemographics-independent preference (GIP). We statistically removed the first two of these components in order to estimate GIP. For the $Hernandez\ conjecture$ to be true, it must be the case that the GIP lies within goldilocks zone; we found this to have small but non-negligible probabilities in 3 sparsely-populated $departamentos$ and zero in the remaining 15; when the country was analyzed as a whole, the probability of the conjecture was zero under several different circumstances.

           **Philip J Gerrish**[1,2]**, Benjamin Zepeda**[3,4]**, T Y Okosun**[5]**, Irene A Gerrish**[6] **& Rosemary Joyce**[7]

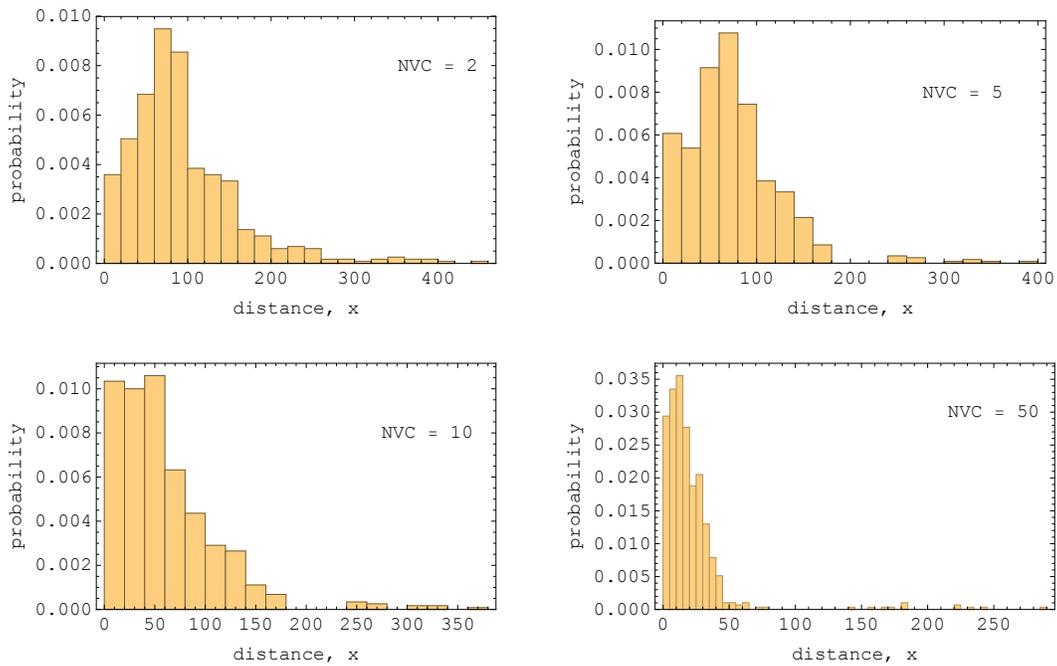

**Fig. S4. Distribution of population density of Honduras as a whole,** $\omega(x)$ (vertical axes), as a function of distance to nearest major voting center, $x$ (horizontal axes). The four panels assume that there are different numbers of major voting centers in the country. To give the incumbent's "victory scenario" argument as much strength as possible, we assumed major voting centers were few and far between. The incumbent's argument relies on effects of distance between voters and voting centers; as such, by looking at cases where these distances are potentially large increases the power of the argument – it is part of the best-case-scenario for the incumbent.

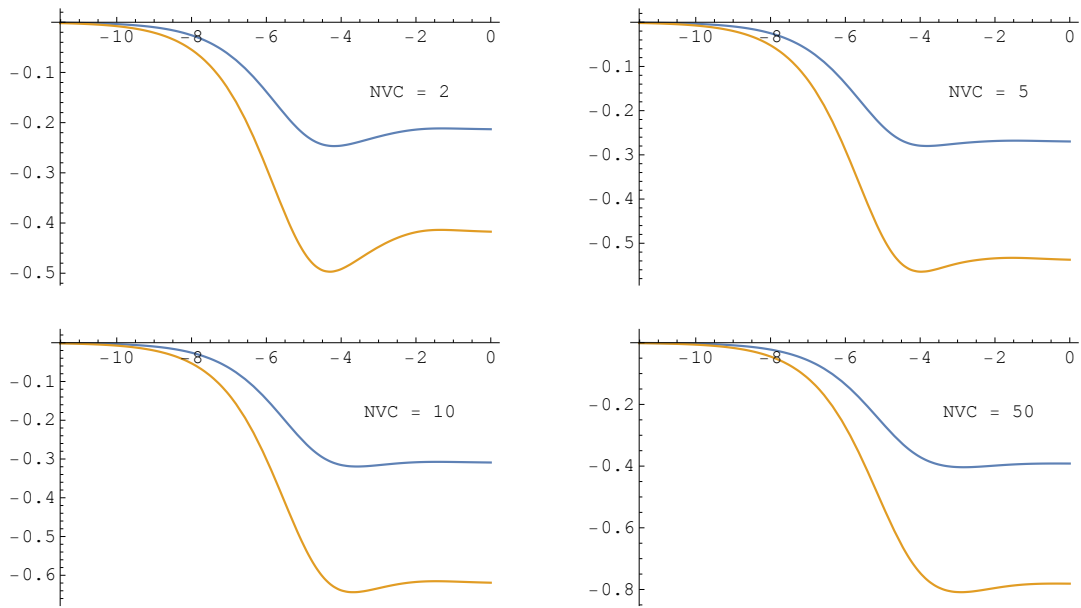

**Fig. S5. The "all-demographics" case: one-parameter exponential model.** We ask the question: assuming we know nothing about voter preferences, can the demographics of Honduras support a "victory scenario" such as the one surmised by the incumbent? From the "no-demographics" case, we know that an ultimate win by the incumbent would require that the Alianza's half-time lead was only an apparent lead – an artifact of demographics and vote-count dynamics – and that, were all votes counted with equal probability at half time, the incumbent would have been in the lead. Mathematically, such a scenario occurs when $\mathbb{E}[g(X)h(X)] < 0$ and $\mathbb{E}[h(X)] > 0$. Blue and orange curves plot $\mathbb{E}[h(X)]$ and $\mathbb{E}[g(X)h(X)]$, respectively, where $X \sim \omega(x)$, against $\log_{10} r$ (horizontal axis), where $r$ is the exponential parameter in the model. The purported victory scenario is thus supported when the orange curve is positive and the blue curve is negative. As evidenced by the plotted curves, we found no case in which this is true.

Philip J Gerrish[1,2], Benjamin Zepeda[3,4], T Y Okosun[5], Irene A Gerrish[6] & Rosemary Joyce[7]

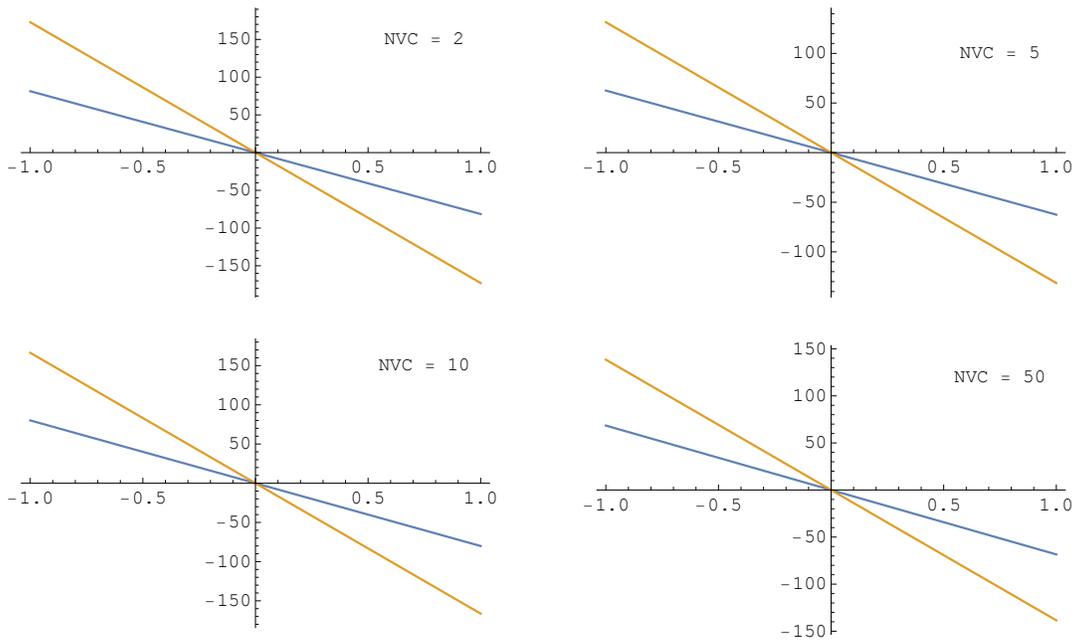

**Fig. S6. The "all-demographics" case: linear model.** We ask the question: assuming we know nothing about voter preferences, can the demographics of Honduras support a "victory scenario" such as the one surmised by the incumbent? From the "no-demographics" case, we know that an ultimate win by the incumbent would require that the Alianza's half-time lead was only an apparent lead – an artifact of demographics and vote-count dynamics – and that, were all votes counted with equal probability at half time, the incumbent would have been in the lead. Mathematically, such a scenario occurs when $\mathbb{E}[g(X)h(X)] < 0$ and $\mathbb{E}[h(X)] > 0$. Blue and orange curves plot $\mathbb{E}[h(X)]$ and $\mathbb{E}[g(X)h(X)]$, respectively, where $X \sim \omega(x)$, against linear slope parameter $m$ (horizontal axis). The purported victory scenario is thus supported when the orange curve is positive and the blue curve is negative. As evidenced by the plotted curves, we found no case in which this is true.

**Philip J Gerrish**[1,2]**, Benjamin Zepeda**[3,4]**, T Y Okosun**[5]**, Irene A Gerrish**[6] **& Rosemary Joyce**[7]

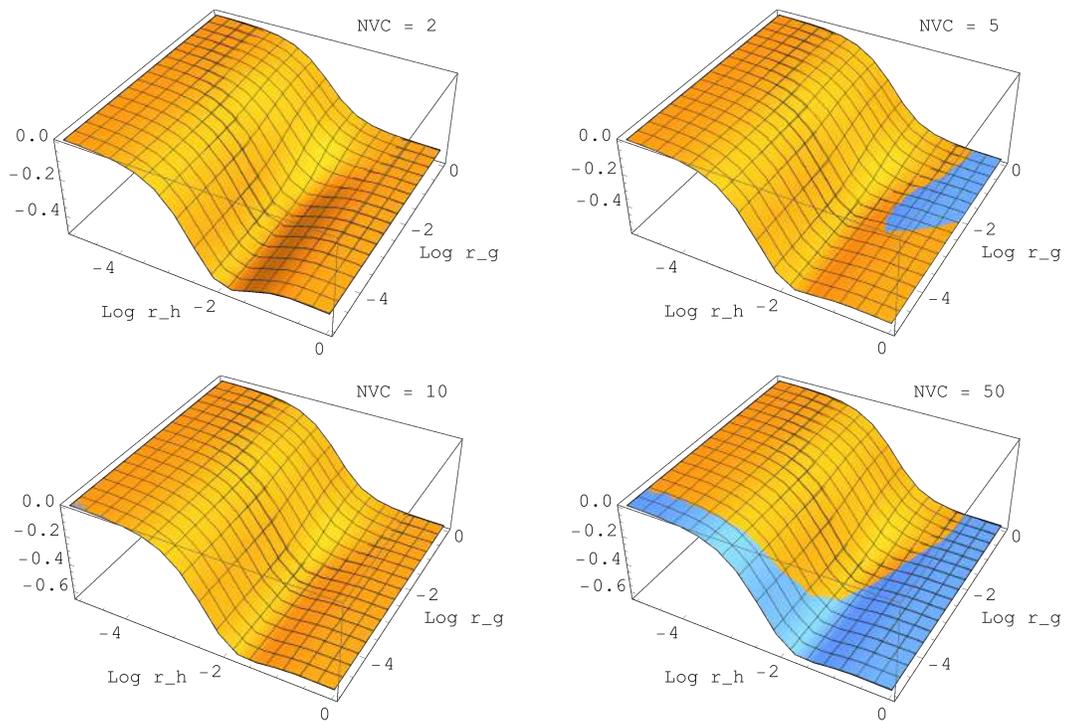

**Fig. S7. The "all-demographics" case: two-parameter exponential model.** We ask the question: assuming we know nothing about voter preferences, can the demographics of Honduras support a "victory scenario" such as the one surmised by the incumbent? From the "no-demographics" case, we know that an ultimate win by the incumbent would require that the Alianza's half-time lead was only an apparent lead – an artifact of demographics and vote-count dynamics – and that, were all votes counted with equal probability at half time, the incumbent would have been in the lead. Mathematically, such a scenario occurs when $\mathbb{E}[g(X)h(X)] < 0$ and $\mathbb{E}[h(X)] > 0$. Blue and orange surfaces plot $\mathbb{E}[h(X)]$ and $\mathbb{E}[g(X)h(X)]$, respectively, where $X \sim \omega(x)$, against $\log_{10} r_h$ (first horizontal axis) and $\log_{10} r_g$ (second horizontal axis), where $r_h$ and $r_g$ are the exponential parameters in the model. The purported victory scenario is thus supported when the orange surface is positive and the blue surface is negative. As evidenced by the plotted surfaces, we found no case in which this is true.

Philip J Gerrish[1,2], Benjamin Zepeda[3,4], T Y Okosun[5], Irene A Gerrish[6] & Rosemary Joyce[7]

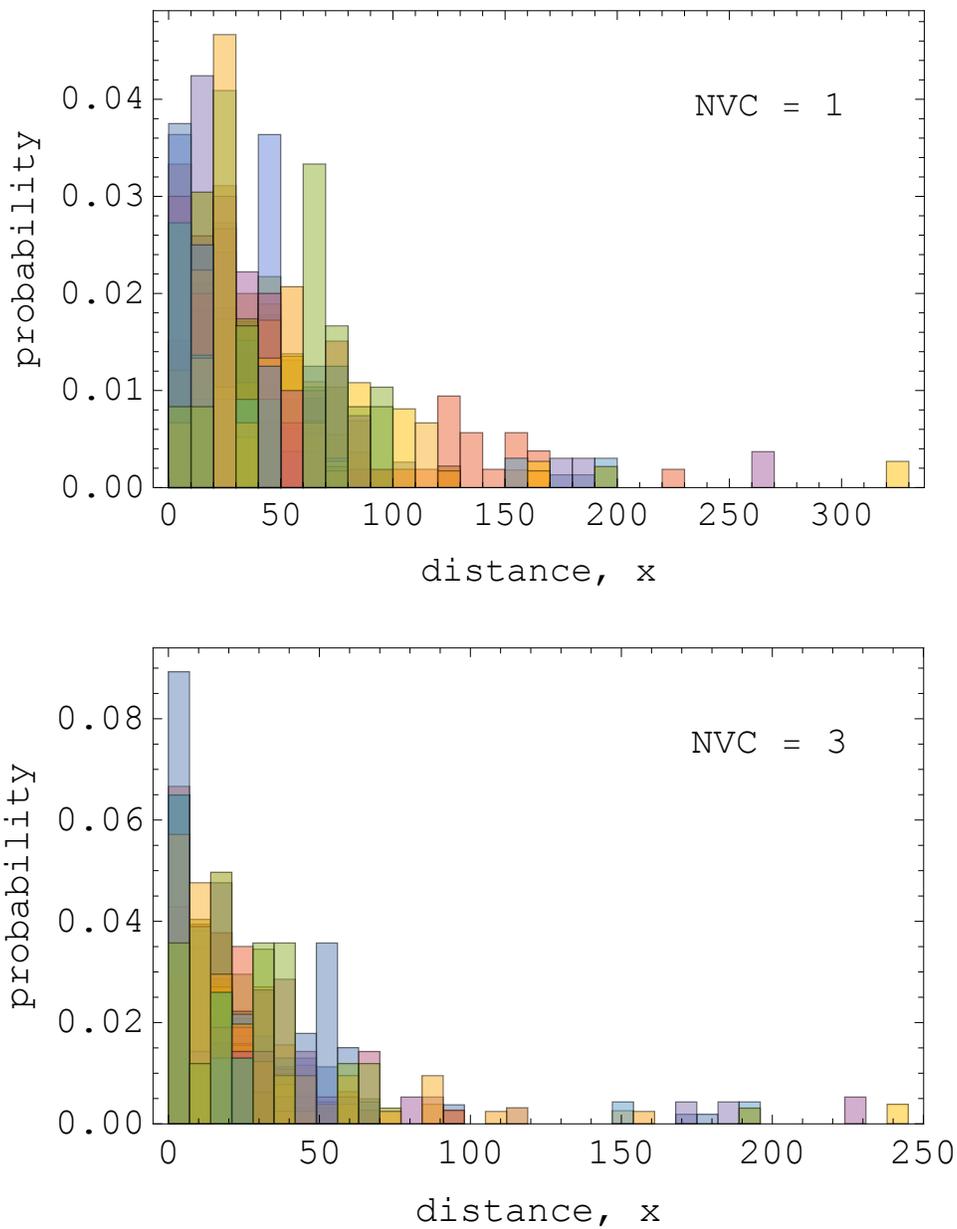

**Fig. S8.** Distribution of population density of Honduras by *departamento*, $\omega(x)$ (vertical axes), as a function of distance to nearest major voting center, $x$ (horizontal axes). A different color is used for each *departamento*. NVC is the number of major voting centers assumed for each *departamento* in the country (NVC=1 and NVC=3). To give the incumbent's "victory scenario" argument as much strength as possible, we assumed major voting centers were few and far between. The incumbent's argument relies on effects of distance between voters and voting centers; as such, by looking at cases where these distances are potentially large increases the power of the argument.

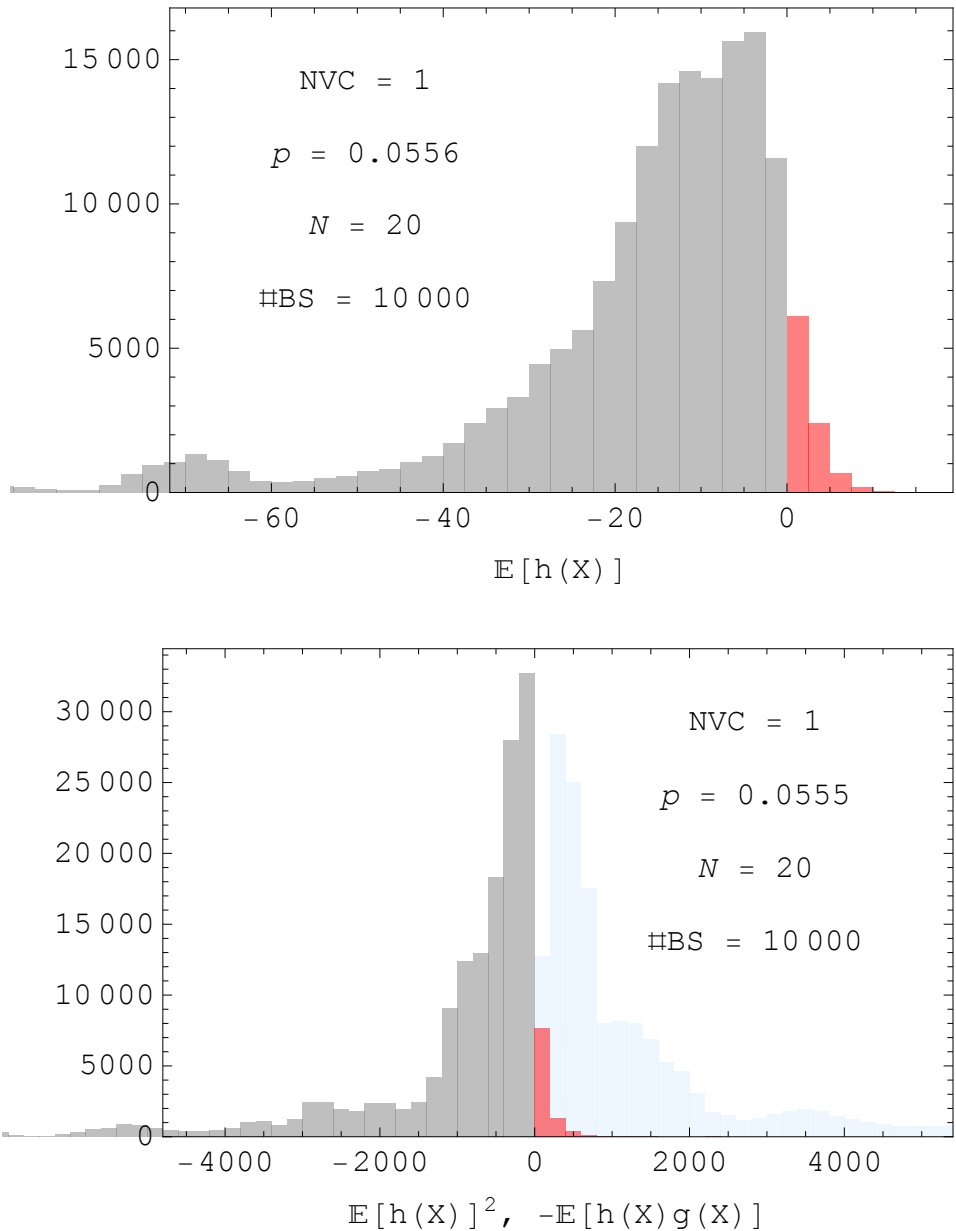

**Fig. S9. A small sample-size look at probabilities of: a) an overall win by Hernandez (top), and b) the *Hernandez conjecture* (bottom).**
Plotted are histograms of the relevant quantities based on jackknifed samples of $X$ from the complete set of geodemographic data of Honduras. The probabilities of interest are indicated by red bars: a) in the top panel, the red bars show the case for which $\mathbb{E}(X) > 0$ indicating an overall win by Hernandez; b) in the bottom panel, the red bars show the overlap between $\mathbb{E}[h(X)]^2$ and $-\mathbb{E}[h(X)g(X)]$, indicating cases for which a ballot turnaround – such as the one observed in the 2017 – is possible. Jackknife sample sizes here are extremely small ($N = 20$) compared to the population of the country ($\sim 8 \times 10^6$), but already the probabilities of a) and b) are small. The number of jackknifed samples is $10,000$ (misleadingly labeled here as "#BS"). $p$-values indicate the fraction of the total counts represented by the red bars. To give the *Hernandez conjecture* as much credibility as possible, we assumed major voting centers were few and far between (here NVC=1 voting center per *departamento*). The incumbent's argument relies on effects of distance between voters and voting centers; as such, by looking at cases where these distances are potentially large increases the power of the argument – it is part of the best-case-scenario for the incumbent.

Philip J Gerrish[1,2], Benjamin Zepeda[3,4], T Y Okosun[5], Irene A Gerrish[6] & Rosemary Joyce[7]

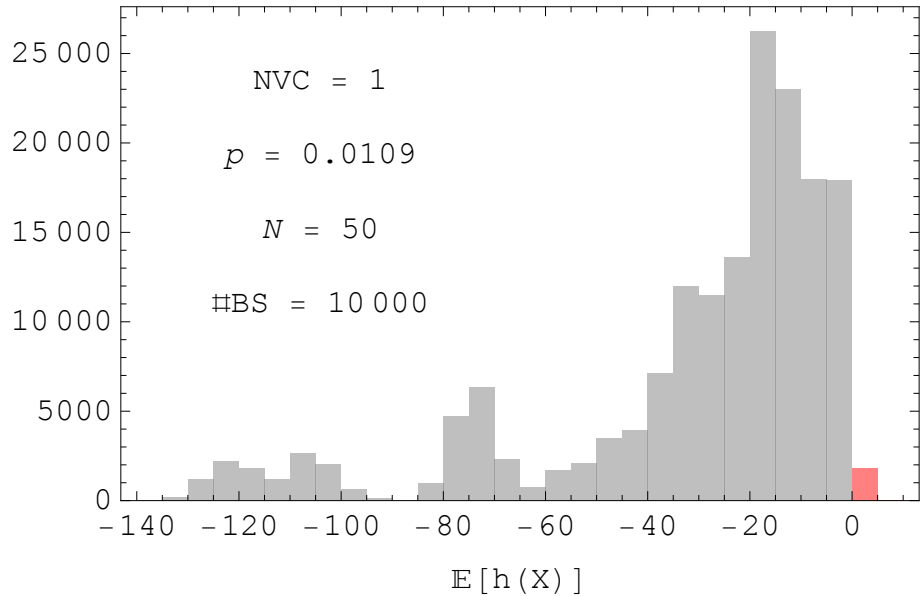

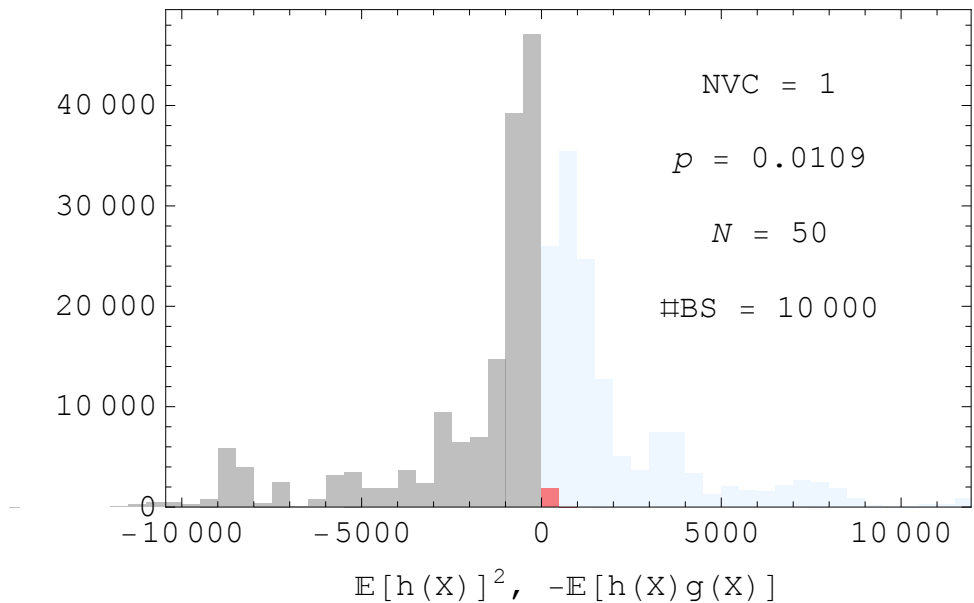

**Fig. S10. A small sample-size look at probabilities of: a) an overall win by Hernandez (top), and b) the *Hernandez conjecture* (bottom).** Plotted are histograms of the relevant quantities based on jackknifed samples of $X$ from the complete set of geodemographic data of Honduras. The probabilities of interest are indicated by red bars: a) in the top panel, the red bars show the case for which $\mathbb{E}(X) > 0$ indicating an overall win by Hernandez; b) in the bottom panel, the red bars show the overlap between $\mathbb{E}[h(X)]^2$ and $-\mathbb{E}[h(X)g(X)]$, indicating cases for which a ballot turnaround – such as the one observed in the 2017 – is possible. Jackknife sample sizes here are extremely small ($N = 50$) compared to the population of the country ($\sim 8 \times 10^6$), but already the probabilities of a) and b) are small. The number of jackknifed samples is $10,000$ (misleadingly labeled here as "#BS"). $p$-values indicate the fraction of the total counts represented by the red bars. To give the *Hernandez conjecture* as much credibility as possible, we assumed major voting centers were few and far between (here NVC=1 voting center per *departamento*). The incumbent's argument relies on effects of distance between voters and voting centers; as such, by looking at cases where these distances are potentially large increases the power of the argument – it is part of the best-case-scenario for the incumbent.

**Philip J Gerrish**[1,2]**, Benjamin Zepeda**[3,4]**, T Y Okosun**[5]**, Irene A Gerrish**[6] **& Rosemary Joyce**[7]

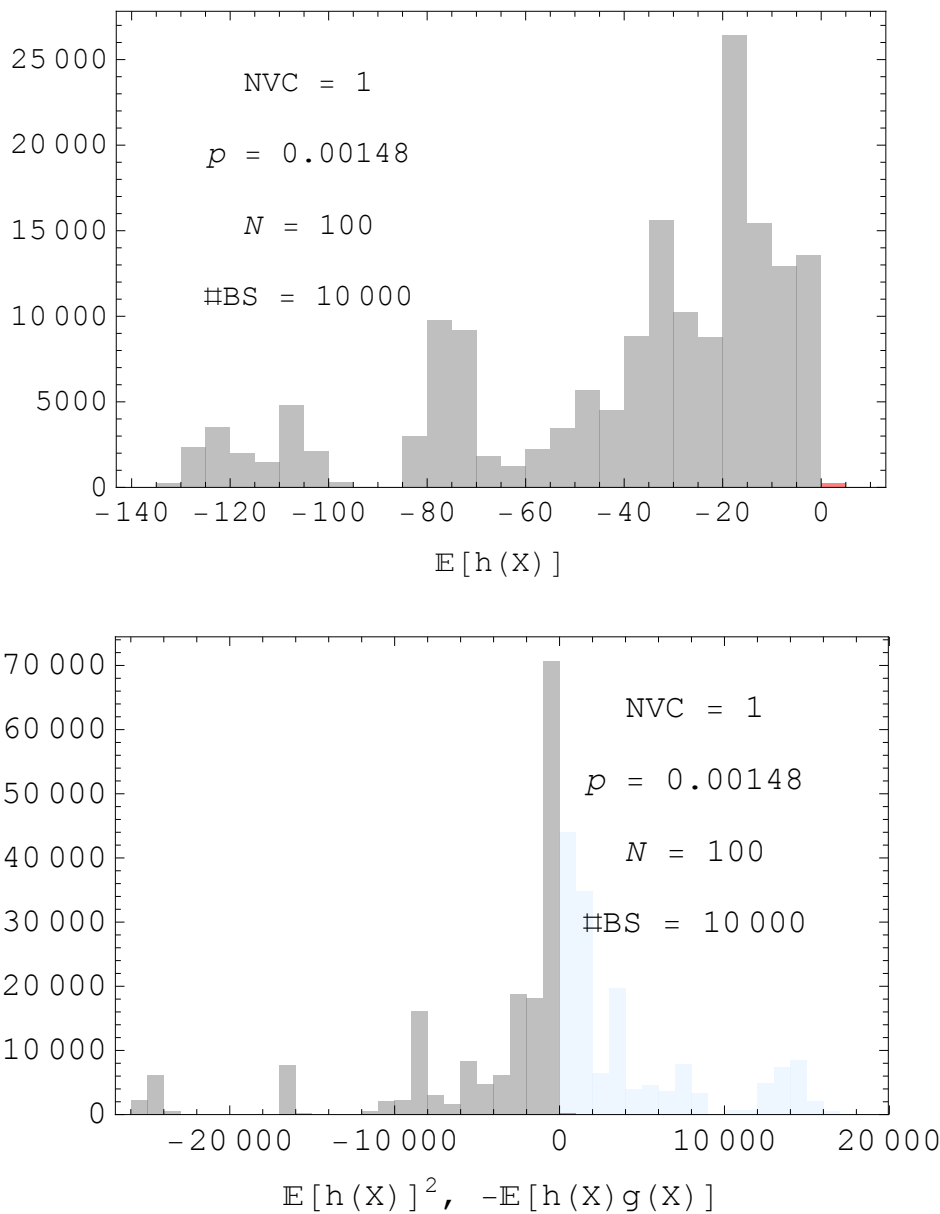

**Fig. S11. A small sample-size look at probabilities of: a) an overall win by Hernandez (top), and b) the *Hernandez conjecture* (bottom).** Plotted are histograms of the relevant quantities based on jackknifed samples of $X$ from the complete set of geodemographic data of Honduras. The probabilities of interest are indicated by red bars: a) in the top panel, the red bars show the case for which $\mathbb{E}(X) > 0$ indicating an overall win by Hernandez; b) in the bottom panel, the red bars show the overlap between $\mathbb{E}[h(X)]^2$ and $-\mathbb{E}[h(X)g(X)]$, indicating cases for which a ballot turnaround – such as the one observed in the 2017 – is possible. Jackknife sample sizes here are extremely small ($N = 100$) compared to the population of the country ($\sim 8 \times 10^6$), but already the probabilities of a) and b) are very small. The number of jackknifed samples is $10,000$ (misleadingly labeled here as "#BS"). $p$-values indicate the fraction of the total counts represented by the red bars. To give the *Hernandez conjecture* as much credibility as possible, we assumed major voting centers were few and far between (here NVC=1 voting center per *departamento*). The incumbent's argument relies on effects of distance between voters and voting centers; as such, by looking at cases where these distances are potentially large increases the power of the argument – it is part of the best-case-scenario for the incumbent.

Philip J Gerrish[1,2], Benjamin Zepeda[3,4], T Y Okosun[5], Irene A Gerrish[6] & Rosemary Joyce[7]

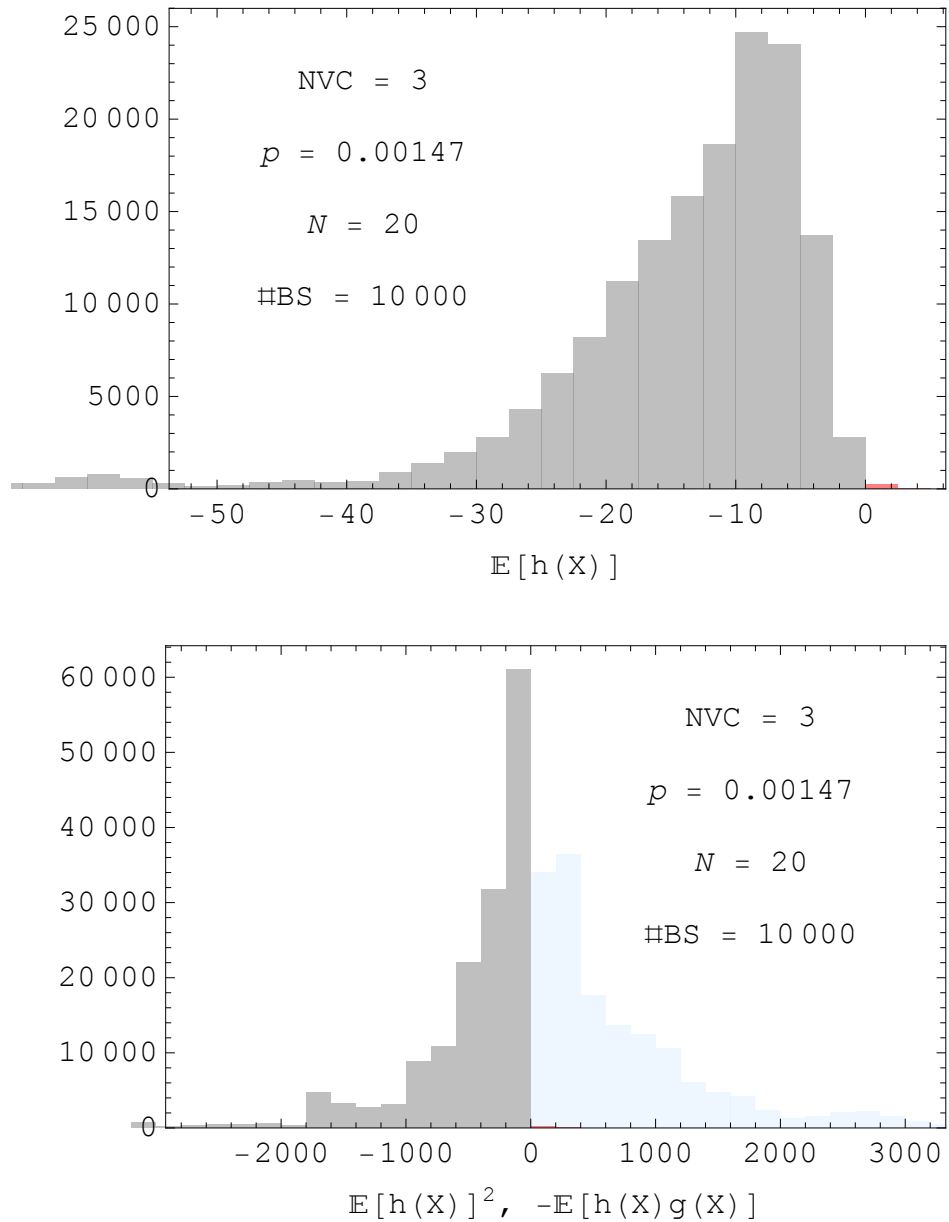

**Fig. S12. A small sample-size look at probabilities of: a) an overall win by Hernandez (top), and b) the *Hernandez conjecture* (bottom).** Plotted are histograms of the relevant quantities based on jackknifed samples of $X$ from the complete set of geodemographic data of Honduras. The probabilities of interest are indicated by red bars: a) in the top panel, the red bars show the case for which $\mathbb{E}(X) > 0$ indicating an overall win by Hernandez; b) in the bottom panel, the red bars show the overlap between $\mathbb{E}[h(X)]^2$ and $-\mathbb{E}[h(X)g(X)]$, indicating cases for which a ballot turnaround – such as the one observed in the 2017 – is possible. Jackknife sample sizes here are extremely small ($N = 20$) compared to the population of the country ($\sim 8 \times 10^6$), but already the probabilities of a) and b) are very small. The number of jackknifed samples is $10,000$ (misleadingly labeled here as "#BS"). $p$-values indicate the fraction of the total counts represented by the red bars. To give the *Hernandez conjecture* as much credibility as possible, we assumed major voting centers were few and far between (here NVC=3 voting centers per *departamento*). The incumbent's argument relies on effects of distance between voters and voting centers; as such, by looking at cases where these distances are potentially large increases the power of the argument – it is part of the best-case-scenario for the incumbent.

**Philip J Gerrish**[1,2], **Benjamin Zepeda**[3,4], **T Y Okosun**[5], **Irene A Gerrish**[6] **& Rosemary Joyce**[7]

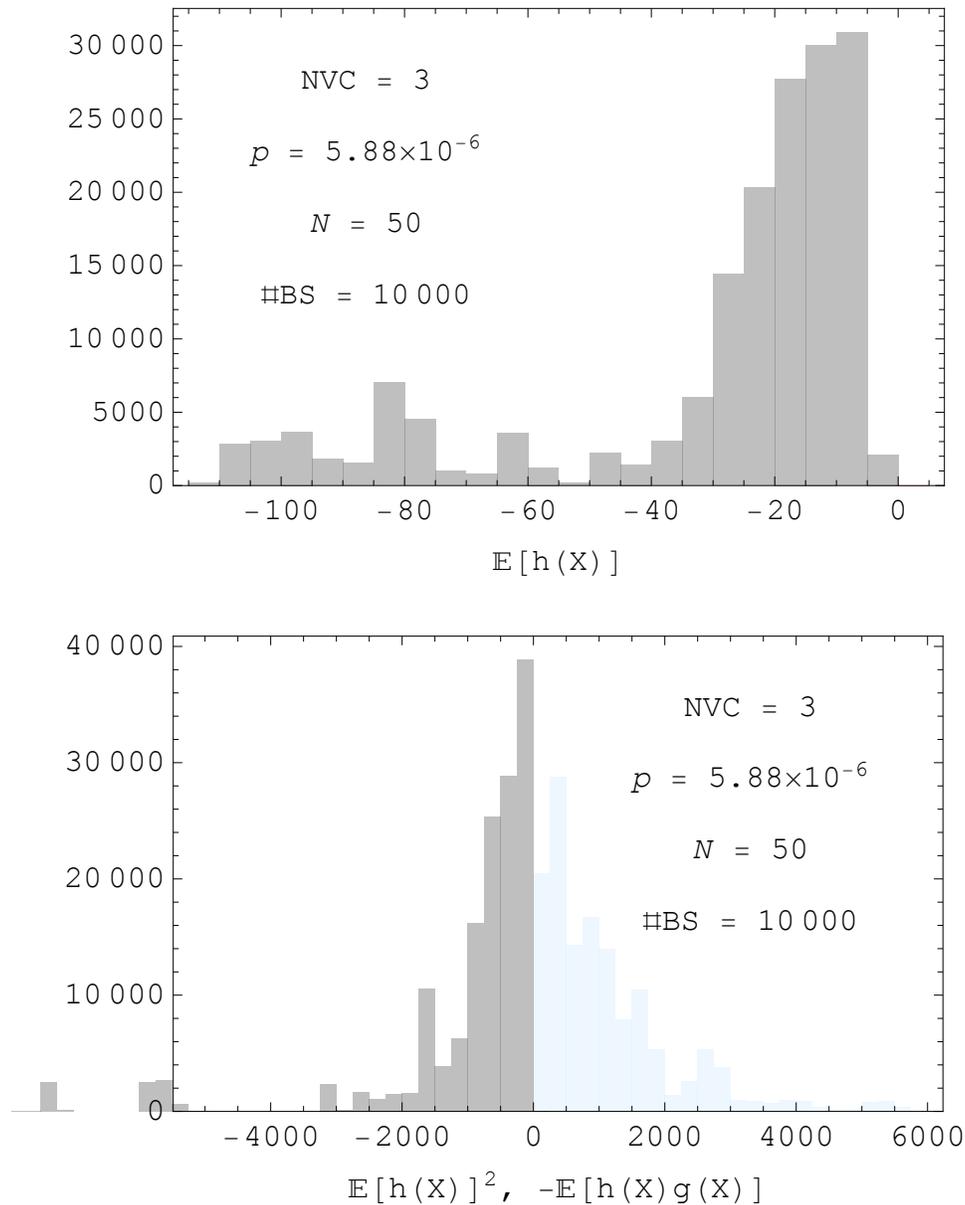

**Fig. S13. A small sample-size look at probabilities of: a) an overall win by Hernandez (top), and b) the *Hernandez conjecture* (bottom).** Plotted are histograms of the relevant quantities based on jackknifed samples of $X$ from the complete set of geodemographic data of Honduras. The probabilities of interest are indicated by red bars: a) in the top panel, the red bars show the case for which $\mathbb{E}(X) > 0$ indicating an overall win by Hernandez; b) in the bottom panel, the red bars show the overlap between $\mathbb{E}[h(X)]^2$ and $-\mathbb{E}[h(X)g(X)]$, indicating cases for which a ballot turnaround – such as the one observed in the 2017 – is possible. Jackknife sample sizes here are extremely small ($N = 50$) compared to the population of the country ($\sim 8 \times 10^6$), but already the probabilities of a) and b) are very small. The number of jackknifed samples is $10,000$ (misleadingly labeled here as "#BS"). $p$-values indicate the fraction of the total counts represented by the red bars. To give the *Hernandez conjecture* as much credibility as possible, we assumed major voting centers were few and far between (here NVC=3 voting centers per *departamento*). The incumbent's argument relies on effects of distance between voters and voting centers; as such, by looking at cases where these distances are potentially large increases the power of the argument – it is part of the best-case-scenario for the incumbent.

   Philip J Gerrish[1,2], Benjamin Zepeda[3,4], T Y Okosun[5], Irene A Gerrish[6] & Rosemary Joyce[7]

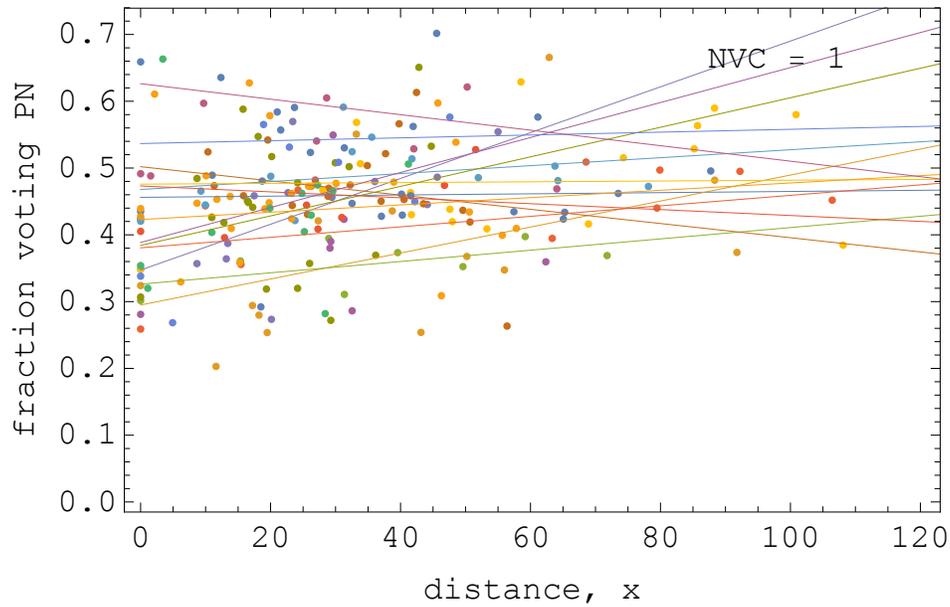

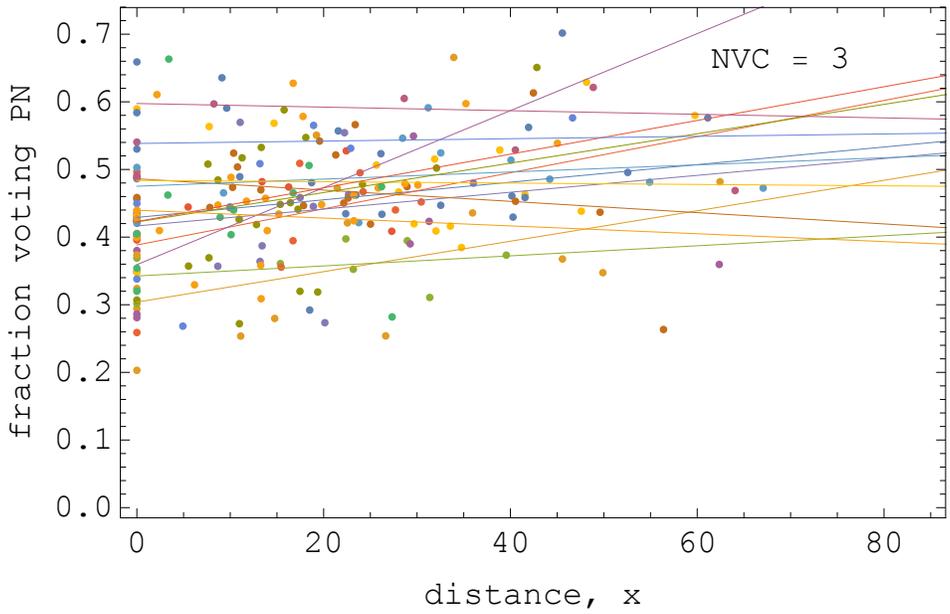

**Fig. S14. Fraction voting for the incumbent (PN) as a function of distance from nearest voting center.** Under the assumption of different numbers of voting centers (NVC). We note that the observed trend runs counter to the trend surmised by the incumbent to explain his victory. Voter data courtesy of The Economist (1). Demographic data of Honduras drawn from Wolfram databases (2).

Philip J Gerrish[1,2], Benjamin Zepeda[3,4], T Y Okosun[5], Irene A Gerrish[6] & Rosemary Joyce[7]

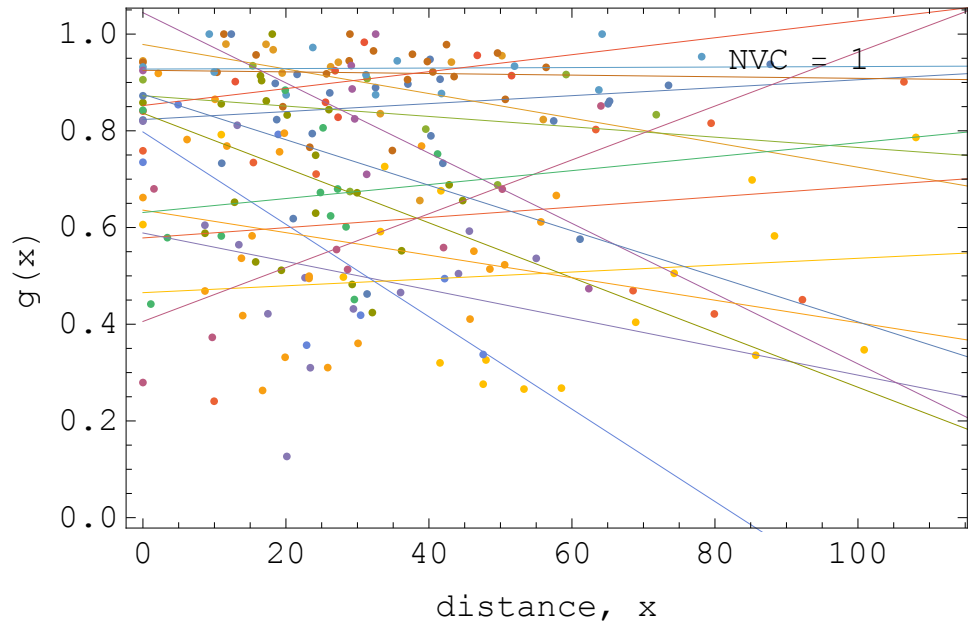

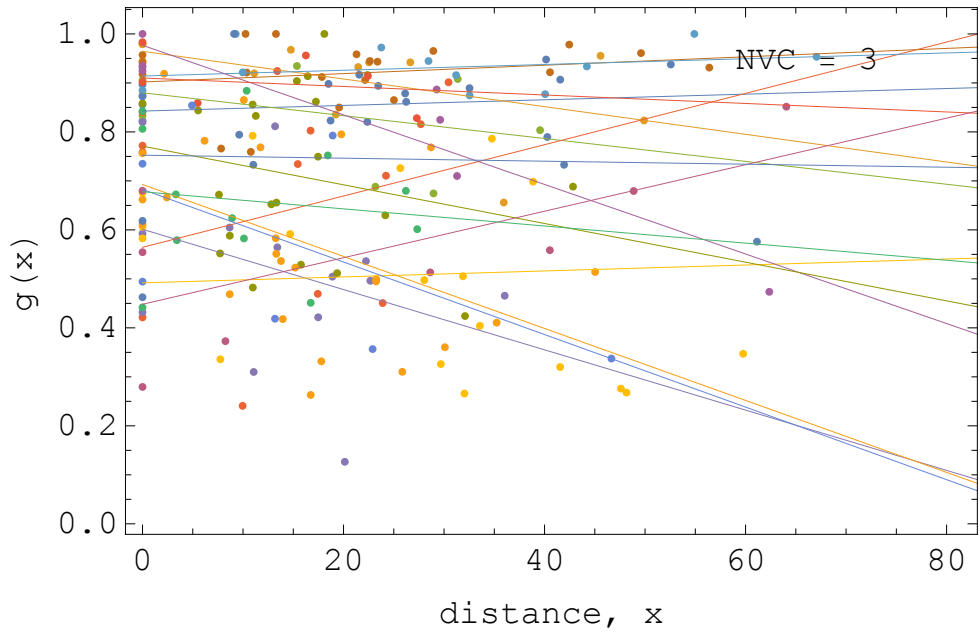

**Fig. S15. Fraction voting for the incumbent (PN) as a function of distance from nearest voting center.** Under the assumption of different numbers of voting centers (NVC). We note that the observed trend runs counter to the trend surmised by the incumbent to explain his victory. Voter data courtesy of The Economist (1). Demographic data of Honduras drawn from Wolfram databases (2).

Philip J Gerrish[1,2], Benjamin Zepeda[3,4], T Y Okosun[5], Irene A Gerrish[6] & Rosemary Joyce[7]

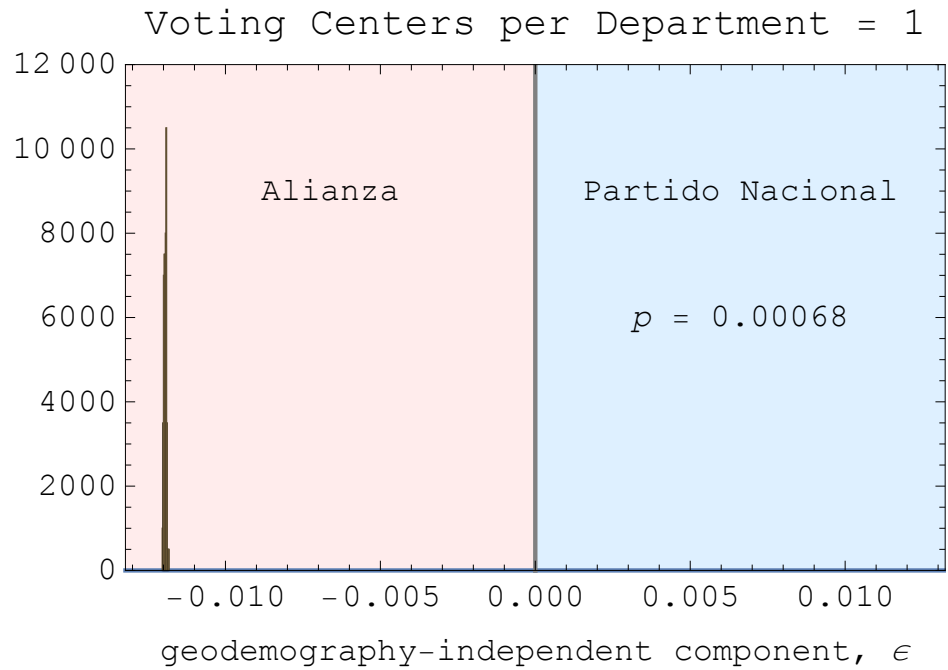

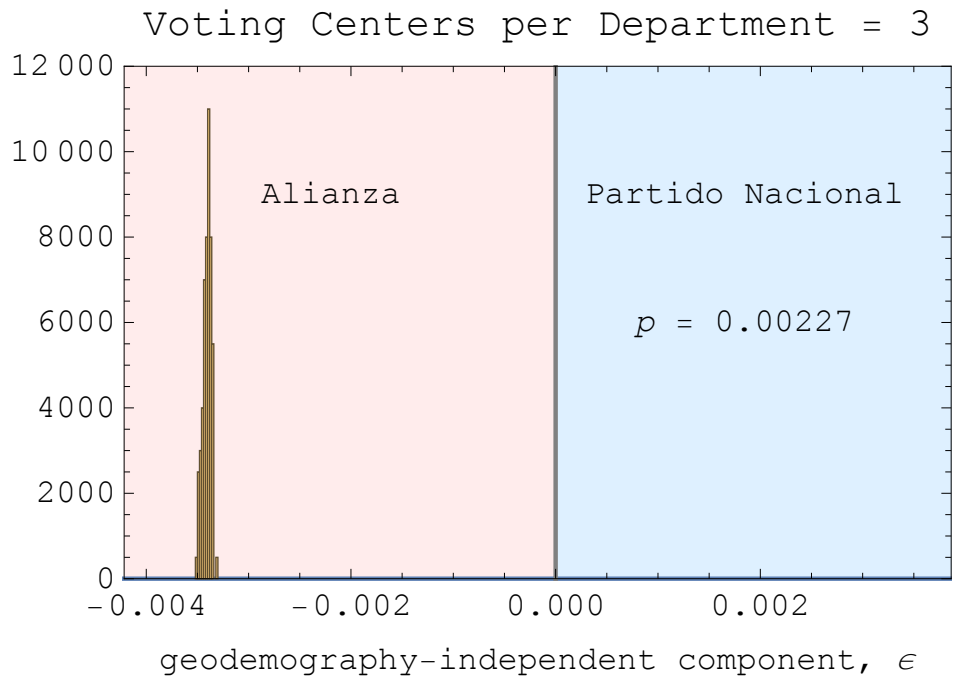

**Fig. S16. Fraction voting for the incumbent (PN) as a function of distance from nearest voting center.** Under the assumption of different numbers of voting centers (NVC). We note that the observed trend runs counter to the trend surmised by the incumbent to explain his victory. Voter data courtesy of The Economist (1). Demographic data of Honduras drawn from Wolfram databases (2).
Philip J Gerrish[1,2], Benjamin Zepeda[3,4], T Y Okosun[5], Irene A Gerrish[6] & Rosemary Joyce[7]

23 of 23